\pdfoutput=1

\documentclass[a4paper, 11pt, abstract=true, titlepage, oneside]{scrartcl}
\makeatletter

\usepackage[T1]{fontenc}
\usepackage[utf8]{inputenc}
\usepackage[onehalfspacing]{setspace}
\usepackage[headsepline]{scrlayer-scrpage}
\usepackage{layout}
\usepackage{geometry}
\usepackage{adjustbox}
\usepackage{authblk}
\usepackage[titletoc,title]{appendix}
\usepackage[hyphens]{url}
\usepackage{color}

\usepackage{amsmath,amssymb,amsfonts,amsthm, mathtools, bm}
\usepackage{mathrsfs}
\usepackage{accents}
\usepackage{thmtools, thm-restate}
\usepackage{nicefrac}
\usepackage{array}

\usepackage[normal]{threeparttable}
\usepackage{threeparttablex}
\usepackage{array,tabularx}
\usepackage{longtable}
\usepackage{booktabs}
\usepackage{colortbl}
\usepackage{multirow}
\usepackage{makecell}
\usepackage{siunitx}

\usepackage{algorithm,algpseudocode}
\usepackage[bottom,hang,multiple]{footmisc}
\usepackage[acronym]{glossaries}

\usepackage{subcaption}
\usepackage{import}
\usepackage{soul}
\usepackage{graphicx}
\usepackage{xcolor} 

\usepackage{placeins}

\usepackage{tikz}
\usetikzlibrary{arrows.meta}
\usetikzlibrary{math}
\usetikzlibrary{shapes}
\usetikzlibrary{patterns,shapes.arrows}
\usepackage{pgfplots}
\pgfplotsset{compat=1.18} 
\usepgfplotslibrary{groupplots}
\usepgfplotslibrary{groupplots,dateplot}
\pgfplotsset{compat=newest}
\usepgfplotslibrary{dateplot}
\usetikzlibrary{backgrounds}
\pgfdeclarelayer{background}
\pgfdeclarelayer{foreground}
\pgfsetlayers{background,main,foreground} 
\usepackage{tikzscale}

\usepackage{enumerate}
\usepackage{multicol}
\usepackage[author=Patrick]{pdfcomment}
\usepackage{xparse}
\usepackage{twoopt}
\usepackage{etoolbox}

\usepackage{eurosym}
\usepackage{ellipsis}
\usepackage{natbib}
\usepackage{paralist}
\usepackage{ulem}
\AtBeginDocument{
	\newgeometry{
		textheight=9in,
		textwidth=6.5in,
		top=1in,
		headheight=16.49675pt,
		headsep=25pt,
		footskip=30pt
	}
}
\newsavebox{\abstractbox}
\renewenvironment{abstract}
{\begin{lrbox}{0}\begin{minipage}{\textwidth}
			\begin{center}\normalfont\sectfont\abstractname\end{center}\quotation}
		{\endquotation\end{minipage}\end{lrbox}%
	\global\setbox\abstractbox=\box0 }

\makeatletter
\expandafter\patchcmd\csname\string\maketitle\endcsname
{\vskip\z@\@plus3fill}
{\vskip\z@\@plus2fill\box\abstractbox\vskip\z@\@plus1fill}
{}{}
\makeatother

\allowdisplaybreaks

\addtokomafont{captionlabel}{\footnotesize\bfseries} 
\addtokomafont{caption}{\footnotesize\bfseries} 

\bibpunct[, ]{(}{)}{,}{a}{}{,}%
\def\newblock{\ }%
\newtheoremstyle{upright} 
  {3pt}                    
  {3pt}                    
  {\upshape}               
  {}                       
  {\bfseries}              
  {.}                      
  {.5em}                   
  {}                       

\theoremstyle{upright}

\let\theoremstyle\relax

\counterwithin*{equation}{section}

\newenvironment{myfont}{\fontfamily{ccr}\selectfont}{\par}
\DeclareTextFontCommand{\textmyfont}{\myfont}

\AtBeginDocument{%
	\abovedisplayskip=7.5pt plus 0pt minus 0pt
	\abovedisplayshortskip=7.5pt plus 0pt minus 0pt
	\belowdisplayskip=7.5pt plus 0pt minus 0pt
	\belowdisplayshortskip=7.5pt plus 0pt minus 0pt
}

\newcolumntype{L}[1]{>{\raggedright\let\newline\\\arraybackslash\hspace{0pt}}p{#1}}
\newcolumntype{C}[1]{>{\centering\let\newline\\\arraybackslash\hspace{0pt}}p{#1}}
\newcolumntype{R}[1]{>{\raggedleft\let\newline\\\arraybackslash\hspace{0pt}}p{#1}}
\renewcommand{\emph}[1]{\textit{#1}}

\algtext*{EndIf} 
\algtext*{EndWhile} 
\algtext*{EndFor} 

\begin{document}
\emergencystretch 3em
\newacronym{acr:mdp}{MDP}{Markov decision process}

\newacronym{acr:ai}{AI}{Artificial Intelligence}

\newacronym{acr:ftc}{FTC}{Federal Trade Commission}

\newacronym{acr:ec}{EC}{European Commission}

\newacronym{acr:rl}{RL}{Reinforcement Learning}

\newacronym{acr:drl}{DRL}{Deep Reinforcement Learning}

\newacronym{acr:ppo}{PPO}{Proximal Policy Optimization}

\newacronym{acr:ppoc}{PPO-C}{\gls{acr:ppo} with continuous action space}

\newacronym{acr:ppod}{PPO-D}{\gls{acr:ppo} with discrete action space}

\newacronym{acr:dqn}{DQN}{Deep Q-Networks}

\newacronym{acr:tql}{TQL}{Tabular Q-learning}

\newacronym{acr:sac}{SAC}{Soft Actor-Critic}

\newacronym{acr:b2c}{B2C}{Business-to-Consumer}

\newacronym{acr:rpdi}{RPDI}{Relative Price Deviation Index}

\title{\large Algorithmic Collusion in Dynamic Pricing with Deep Reinforcement Learning}

\author[1]{\normalsize Shidi Deng}
\author[2]{\normalsize Maximilian Schiffer}
\author[3]{\normalsize Martin Bichler}
\affil{\small 
	School of Management, Technical University of Munich, Germany	
 
	\scriptsize shidi.deng@tum.de

        \small
	\textsuperscript{2}School of Management \& Munich Data Science Institute,
	
	Technical University of Munich, Germany
	
	\scriptsize schiffer@tum.de

         \small
	\textsuperscript{3}School of Computation, Information and Technology, 
      Technical University of Munich, Germany
	
	\scriptsize bichler@cit.tum.de

 }

\date{}

\lehead{\pagemark}
\rohead{\pagemark}

\begin{abstract}
\begin{singlespace}
{\small\noindent Nowadays, a significant share of the \glsentrylong{acr:b2c} sector is based on online platforms like Amazon and Alibaba and uses \glsentrylong{acr:ai} for pricing strategies. This has sparked debate on whether pricing algorithms may tacitly collude to set supra-competitive prices without being explicitly designed to do so. Our study addresses these concerns by examining the risk of collusion when \glsentrylong{acr:rl} algorithms are used to decide on pricing strategies in competitive markets. Prior research in this field focused on \gls{acr:tql} and led to opposing views on whether learning-based algorithms can lead to supra-competitive prices. Our work contributes to this ongoing discussion by providing a more nuanced numerical study that goes beyond \gls{acr:tql} by additionally capturing off- and on-policy \gls{acr:drl} algorithms. We study multiple Bertrand oligopoly variants and show that algorithmic collusion depends on the algorithm used. In our experiments, \gls{acr:tql} exhibits higher collusion and price dispersion phenomena compared to \gls{acr:drl} algorithms. We show that the severity of collusion depends not only on the algorithm used but also on the characteristics of the market environment. We further find that \glsentrylong{acr:ppo} appears to be less sensitive to collusive outcomes compared to other state-of-the-art \gls{acr:drl} algorithms. \\


\smallskip}
\end{singlespace}
\end{abstract}

\maketitle
\section{Introduction}
\label{sec:introduction}
Over the past two decades, many \gls{acr:b2c} businesses have shifted from traditional to online platforms such as Amazon and Alibaba. With the advent of \gls{acr:ai} technology and big data, platform sellers tend to rely progressively on pricing algorithms to explore market dynamics and demand elasticity. Algorithms used in this context often learn pricing strategies via self-play, i.e., without prior knowledge and with little guidance from their developers, as it is the case when using \gls{acr:rl}. Against this background, scientists, institutions such as the \gls{acr:ftc} and the \gls{acr:ec}, as well as practitioners started to debate whether such algorithms may tacitly collude in today's markets \citep{constantine2018oecd, capobianco2020competition}. Algorithmic collusion describes a phenomenon, where independent pricing algorithms learn to set supra-competitive prices higher than the Nash equilibrium. This collusion is tacit in the sense that the algorithms are not programmed to collude explicitly.

So far, there is no consensus on whether algorithmic collusion exists and whether it leads to supra-competitive prices in today's markets and as such impacts social welfare. This missing consensus reflects the fact that analyzing implicit collusive behavior among algorithms poses theoretical and practical challenges. On the one hand, analyzing real-world markets suffers from clear measures of algorithmic collusion and is often additionally impaired by missing data or information due to confidentiality reasons. On the other hand, analyzing algorithmic collusion in stylized market models that allow the detection of tacit collusion via comparison to a closed-form equilibrium, is often criticized for being too narrowly focused on a single market model or a specific algorithm.

Recently, many works evolved that focus on simulated market environments to test the behavior of different pricing algorithms \citep{sanchez2022artificial, kastius2022dynamic, asker2022artificial}, usually focusing on \gls{acr:rl} algorithms which are widely used in commercial pricing scenarios \citep{charpentier2021reinforcement, mosavi2020comprehensive}. Of course, these models are an abstraction and simplification where the same pricing game is played repeatedly by the same sellers. Reality is more complex and there are changes in supply and demand. However, if we observe algorithmic collusion already in this model environment, it might well be a concern in real-world markets.
Still, most of these works focus on a single market model and a basic \gls{acr:rl} algorithm: \gls{acr:tql} \citep{calvano2020artificial,klein2021autonomous}. Against this background, we aim to provide a more nuanced numerical study that captures \gls{acr:drl} algorithms beyond \gls{acr:tql} in order to shed light on the risk of collusion in the presence of algorithmic pricing.

\paragraph{State of the art}
Our work relates to the growing literature on algorithmic pricing. Different researchers have primarily analyzed the role of \gls{acr:ai} algorithms in pricing via numerical studies. For example, \citet{klein2021autonomous} analyzed how \gls{acr:tql} algorithms achieve human-like collusive behavior in a sequential pricing environment. \citet{calvano2020artificial} found that the \gls{acr:tql} algorithm in Bertrand's model with logit demand tends to employ collusive strategies to achieve supra-competitive prices and maintain high prices through time-limited penalties. \citet{sanchez2022artificial} considered the competing behavior of Q-learning in three different market structures. \citet{asker2021artificial} examined the influence of \gls{acr:ai} learning protocols on pricing outcomes in a simple Bertrand game. However, the main limitations of these studies are their reliance on basic \gls{acr:rl} algorithms such as \gls{acr:tql} and their application to a single economic modeling context. It remains an open question whether collusion arises for different \gls{acr:rl} algorithms and in different versions of the Bertrand oligopoly model.

While algorithmic collusion has been widely discussed in simulations, empirical studies on its impact in practice are rare. Among these scarce studies, \citet{assad2020algorithmic} and \citet{brown2021competition} found that algorithmic pricing increased gas station profits and retail drug prices, while also widening price disparities. In recent years, amid vigorous debates on algorithmic collusion, extensive research has delved into how collusion might impact antitrust regulations, necessitating prompt policy responses \citep{werner2023algorithmic,constantine2018oecd, capobianco2020competition}. 

Although the concept of algorithmic pricing collusion and related research evolved recently, the resulting findings and subsequent work in this area have sparked intense academic debate and discussion. On the one hand, the studies by  \citet{meylahn2022learning} and \citet{asker2022artificial} criticized the methodological basis and experimental design of these studies, pointing out possible theoretical and practical problems. On the other hand, the work of \citet{epivent2022algorithmic} questions the interpretation and conclusions of these findings, emphasizing the need for more profound studies and comprehensive analyses to validate these findings. 

\paragraph{Contribution}
With this work, we aim to contribute to a more profound understanding of algorithmic collusion for \gls{acr:rl} and in particular \gls{acr:drl}, which are increasingly used for pricing and bidding by companies \citep{zhou2022deep, afshar2022automated, liu2019dynamic}. 
We provide a comprehensive analysis of tacit collusion in the context of algorithmic pricing via \gls{acr:rl}-based algorithms. 
Contrary to existing works, we do not focus on a single algorithm but cover, besides the widely studied \gls{acr:tql} algorithm, state-of-the-art  \gls{acr:drl} algorithms such as \gls{acr:dqn}, \gls{acr:sac}, and \gls{acr:ppo}. We further provide more nuanced numerical evidence by studying a Bertrand competition in its standard variant but also under different demand models and with varying constraints on production capacity and product heterogeneity, i.e., a Bertrand-Edgeworth competition and a Bertrand competition with logit demand.

We present an extensive numerical study that partially confirms the findings of recent research but puts these into a different perspective, by unraveling new dynamics and insights when extending analyses beyond a standard Bertrand competition and towards more sophisticated \gls{acr:drl} algorithms. We confirm that algorithmic collusion is a concern with \gls{acr:tql} independent of the underlying market model but find different effects when employing \gls{acr:drl} algorithms: here, we observe collusion in all cases when analyzing dynamics in the Bertrand Edgeworth model, while the Logit Bertrand model leads to both scenarios in which we observe collusion and scenarios in which we observe competitive outcomes. In the standard Bertrand model, only one \gls{acr:ppo} variant reaches competitive and collusive outcomes, while other algorithms always collude or show dispersion effects. We report dispersion effects, i.e., two algorithms converging at different prices, mainly for \gls{acr:tql} and \gls{acr:dqn}. Such effects remain artificial artifacts that result from poor exploration during learning. In general, our results suggest that collusion is impacted by an algorithms exploration scheme, which is why on-policy algorithms that implicitly explore during learning lead to less collusion compared to the tested off-policy algorithms and \gls{acr:tql}.

\section{Methodology}
\label{sec: methodologies}
In the following, we first detail the studied market model and its respective variants, before elaborating on our pricing framework, and introducing the studied \gls{acr:rl} algorithms.

\subsection{Market model}
\label{subsec:economic_environment}
We study the Bertrand model \citep{bertrand1883review}, which depicts oligopolistic competition among firms that produce (homogeneous) products and compete by setting prices. In this context, each company uses an independent pricing algorithm.
We limit our study to duopolistic competition, i.e., a market with two companies, labeled as $i=0,1$. While this appears limiting from a game theoretical perspective, it allows us to isolate algorithm dynamics between two players without inheriting numerical instabilities that evolve in multi-player settings. 
Each company $i$ produces a product of a certain quality $g$ and incurs a corresponding marginal cost $c$. The Bertrand model primarily examines the interaction between companies setting product prices $p_i$ and consumers, who choose offers based on these prices, directly affecting the demand $d_i$ for each company's product. In this setting, company $i$'s profit reads $\pi_i = (p_i - c)\times d_i$. And companies set prices simultaneously, aiming to maximize their profit. The Nash equilibrium price, $p^N$, is a price vector where none of the companies has an incentive for unilateral deviation. The monopoly price, $p^M$, is where we treat all companies as a single entity that maximizes profits without competition. This interaction between companies and consumers leads to an interplay between price and demand which is repeated over multiple rounds. At the start of each round, the two companies simultaneously decide on the prices for their respective products. Based on these prices, consumers choose their demand for each product. Afterward, the interaction between firms progresses to the next round. 
We study different variants of the Bertrand model, which vary with respect to their assumptions on product capacity, product quality, and the demand function.

In the \textbf{standard Bertrand competition}, products are homogeneous, and consumers choose the less expensive products. The demand function reads $d(p)=1-p$, and given prices $p_i,p_{-i}$ of both companies, the demand splits according to  
    \begin{align}
        d_i(p_i,p_{-i}) = \begin{cases} 
        d(p_i) & \text{if } p_i<p_{-i},  \\
        \frac{1}{2}d(p_i) & \text{if } p_i= p_{-i}, \\
        0 & \text{if } p_i>p_{-i}
        \end{cases},
    \end{align}
i.e., due to infinite product capacity on the company's side, either the company with the lower price captures all demand or equal prices result in an equal demand split.

In the \textbf{Bertrand-Edgeworth competition}, a company's supply is limited by its production capacity \citep{edgeworth1925papers}, which can lead to market prices higher than the marginal costs in case capacity limits are met. While this adds a realistic constraint to the standard Bertrand competition, it complicates closed-form equilibrium analyses.  To ensure a unique and well-defined Nash and monopoly price, we focus on a special case, where two competing companies have the same production capacity $k > 0.5$, their pricing strategies are constrained to the interval $[0, 1]$, and the demand function reads $d(p) = 1-p$. Then, the demand splits according to
\begin{align}
d_i(p_i,p_{-i}) = \begin{cases} 
\min\{k, 1-p_i\} & \text{if } p_i < p_{-i}, \\
\frac{1-p_i}{2} & \text{if } p_i = p_{-i}, \\
\max\{0,  1-p_i-k\} & \text{if } p_i > p_{-i}
\end{cases}, 
\end{align}
and the Nash and monopoly results remain consistent with the standard Bertrand model as the total capacity of the two firms exceeds the market demand for $p_0=p_1=c$.

In the \textbf{Bertrand's model with logit demand}, the demand for a company $i$'s product is given by:
\begin{equation}
   d_i(p_i, p_{-i})=\frac{e^{\frac{g-p_{i}}{\mu}}}{\sum_{j=0}^1 e^{\frac{g-p_{j}}{\mu}}+1}. 
\end{equation}
Here $g-p_i$ represents the utility that consumers derive from purchasing the product $i$. The parameter $\mu$ captures inter-product substitutability, which is higher when $\mu$ is lower.


\subsection{Algorithmic framework}
We study the dynamics of algorithmic collusion by \gls{acr:drl}-based agents, each representing a company, that interact via pricing decisions, aiming to maximize profit. In this context, we formalize an agent's decision-making as an infinite time horizon \gls{acr:mdp} with decision times $t\in\{1,\ldots,\infty\}$, reflecting the continuous interaction and competition among companies in the market. At each time step $t$, agents simultaneously take actions by deciding on a price $p_t\in \mathcal{A}$ that lies within their action space $\mathcal{A}$. When taking this action, agents take the current state $s_t \in \mathcal{S}$ into account. Here, a state is a tuple that contains each agent's pricing decision from the previous time step such that $s_t = (p_{i,t-1})_{i = 0,1}$. 
The action space $\mathcal{A}$ for the agents is constrained by a possible price range $[\underline{p}, \bar{p}]$ for their products, which we divided into $m$ equidistant values in case we model a discrete action space. Accordingly, the state space $\mathcal{S} = \mathcal{A}\times \mathcal{A}$ formally results in a quadratic growth with an increase in $m$. After agents take their actions at time step $t$, the environment transitions to the next state $S_{t+1}$, and each agent receives a reward $R_t$ based on the respective demand, representing the firm's profits. The agent aims to choose actions $A_t$ to maximize its discounted value of future rewards $G_t$, where $G_t=\sum_{k=0}^{\infty}\gamma^{k}R_{t+k+1}$ denotes discounting future rewards, reflecting the current valuation of future profits. 

Specifically, in the standard Bertrand and Bertrand-Edgeworth models, we set the price range between $[0, 1]$. For the Bertrand model with logit demand, we follow the definitions in \citep{calvano2020artificial}, setting $\underline{p} = p^N - \zeta(p^M - p^N)$ and $\bar{p} = p^M + \zeta(p^M - p^N)$, where $\zeta$ indicates the flexibility of pricing strategies. 

\subsection{\gls{acr:drl} algorithms}
\label{subsec: rl_algorithms}



In the above \gls{acr:mdp} framework, an agent selects an action according to a policy $\pi(a|s_t)$ at each time step $t$. This policy can be deterministic or stochastic and determines an agent's pricing behavior, i.e., the chosen action $a\in\mathcal{A}$ when being in state $s_t$. In \gls{acr:rl} an agent learns such a policy $\pi$ by evaluating state-action pairs $(s,a)$ using a Q-function that reads $Q(s, a) = E \left[\sum_{k=0}^{\infty} \gamma^k R_{t+k+1} | S_t = s, A_t = a\right]$. A policy $\pi^*$ is optimal if it produces no less expected return for all states $s$ and actions $a$ than any other policy. The essence lies in the optimal action-value function's Bellman equation: $Q^*(s, a) = \mathbb{E} \left[R_{t+1} + \gamma \max_{a'} Q^*(s', a') | S_t = s, A_t = a\right]$. This indicates that the maximum expected reward for taking action $a$ in state $s$ and following the optimal strategy is the immediate reward plus the expected maximum action value in the next state. 

The core goal of all \gls{acr:rl} algorithms is to learn such an optimal policy to maximize the long-term rewards. Still, these algorithms can differ in the algorithmic approach used to learn $\pi$. In this work, we consider the following algorithms to account for different learning dynamics.

\textbf{\gls{acr:tql}}, introduced by \citet{watkins1989learning}, is an off-policy algorithm, capable of learning an optimal policy that is different from the policy it follows during exploration. This method approximates the respective  Q-function in a tabular form, i.e.,  within a matrix $Q_0$ of dimensions $|S|\times|A|$, where $|S|$ and $|A|$ represent the size of the state and action spaces, respectively. These values are asynchronously updated at each step of the agent's interaction with the environment. The update rule utilizes the Bellman equation: $Q_{t+1}(s, a) =(1-\alpha)Q_t(s,a)+\alpha[R_{t+1} + \gamma \max_{a'} Q_t(s', a')]$, utilizing a temporal-difference update where $\alpha$ is the learning rate and $s'$ is the state at time step $t+1$.

\textbf{\gls{acr:dqn}}, proposed by \citet{mnih2015human}, utilizes deep neural networks with weights $\theta$ to approximate the optimal action-value function $Q^*(s, a)$, denoted as $Q(s, a; \theta)$. This approach enables the handling of complex state-action spaces, overcoming the scalability limitations of \gls{acr:tql}. The use of neural networks allows \gls{acr:dqn} to learn from the experience of state-action pairs $(s,a)$ and apply this knowledge to similar pairs, thereby enhancing learning efficiency via generalization. Furthermore, \gls{acr:dqn} enhances the stability of the learning process by incorporating the concepts of experience replay to decorrelate samples and target networks to stabilize the loss evaluation.

\textbf{\gls{acr:ppo}}, developed by \citet{schulman2017proximal}, is tailored for both continuous and discrete action spaces. Unlike \gls{acr:tql} and \gls{acr:dqn}, which employ off-policy learning and focus on value function approximation, \gls{acr:ppo} parameterizes the policy $\pi(s, a)$, denoted as $\pi(s, a; \theta)$, using deep neural networks. It directly maps observed states to action probability distributions, thus facilitating policy optimization by adjusting the network parameters $\theta$. As an on-policy algorithm, \gls{acr:ppo} enhances sample efficiency and simplifies implementation by learning and improving the same policy responsible for generating actions. Moreover, \gls{acr:ppo} utilizes gradient clipping to stabilize convergence when learning the policy networks parameterization $\theta$ via gradient descent. This limits the extent of policy changes, ensuring that updates are substantial enough to boost performance without leading to detrimental volatility.

\textbf{\gls{acr:sac}}, developed by \cite{haarnoja2018soft}, is an actor-critic method that integrates policy optimization with value function learning, tailored for robust sampling in continuous action spaces. 
\gls{acr:sac} uses policy gradients and soft Q-values to improve sampling efficiency and performance. It estimates the value of state-action pairs by learning the value function and minimizing the approximation error while using a policy's entropy to maintain diverse exploration. Additionally, adaptive temperature parameters help balance rewards and exploration, allowing \gls{acr:sac} to adapt to various environments. Adjustable entropy regularization and soft Q-value parameters ensure stable policy updates, making \gls{acr:sac} perform well in continuous control tasks.

\section{Experimental design}
\label{subsec:experimental_design}
For our numerical studies, we use the following experimental design. We start by running two homogeneous \gls{acr:rl} agents in each setting to mimic two firms with similar characteristics. Both agents have the same hyperparameter settings to ensure fairness. 
If \gls{acr:tql} is used, we set the number of total timesteps $T$ for one run to 2,000,000  to account for its slow-learning nature. Here, we set the corresponding discount factor $\gamma$ to 0.95 and the learning rate to $\alpha = 0.125$. As for the \gls{acr:drl} algorithms, \gls{acr:dqn}, \gls{acr:ppo} and \gls{acr:sac}, agents compete over 200,000 time steps. Their corresponding standard settings then include a discount factor $\gamma$ of 0.99. For \gls{acr:ppo}, the learning rate $\alpha$ is 0.00005; for \gls{acr:dqn} it is 0.0001; and for \gls{acr:sac} it is 0.0003. We refer to Appendix~\ref{appendix:experimental_setup} for a detailed discussion on how we selected hyperparameters for each algorithm. Note that we use a fixed number of time steps in our experiments to accommodate learning algorithms with slow convergence. 
We evenly divide the action intervals to suit the discrete action space. For \gls{acr:tql} and \gls{acr:dqn}, which only support discrete actions, the action space includes $m = 15$ price options. For \gls{acr:ppo}, we consider two variants: \gls{acr:ppoc} and \gls{acr:ppod}. We apply \gls{acr:sac} to continuous action spaces.

We parameterize three variants of the Bertrand competition model to investigate the effects of different market structures on competitive behavior. In the standard Bertrand and Bertrand-Edgeworth models, we set the marginal cost $c$ of all competitive agents to 0. However, in the Bertrand model with logit demand, we set $c$ to 1 to match the data used in \citet{calvano2020artificial}. In the Bertrand-Edgeworth model, we assign a capacity constraint $k$ of 0.6 to both agents. For the Bertrand model with logit demand, we set product quality $g$ to 2 and inter-product substitutability $\mu$ to 0.25. For the standard Bertrand model and the Bertrand-Edgeworth model, we derive the Nash price $p^{N}$ and the monopoly price $p^{M}$ to be 0 and 0.5, respectively, and the Nash profit $\pi^{N}$ and the monopoly profit $\pi^{M}$ to be computed to be 0 and 0.125, respectively. In the Bertrand model with logit demand, the Nash price $p^{N}$ increases to 1.473, the monopoly price $p^{M}$ increases to 1.925, and the Nash profit $\pi^{N}$ and monopoly profit $\pi^{M}$ are calculated as 0.223 and 0.337, respectively. In all experiments, we maintained consistent parameters to ensure uniformity. 
\section{Results}
In the following, we discuss the results of our numerical experiments. We first discuss the observed pricing dynamics, before elaborating on the causes of the observed effects, and discussing limitations and future research perspectives.

\subsection{Pricing dynamics}\label{subsec:pricingDynamics}
\begin{figure}[!hb]
  \centering  
  \begin{subfigure}{0.32\linewidth}
    \centering
    \includegraphics[width=\linewidth]{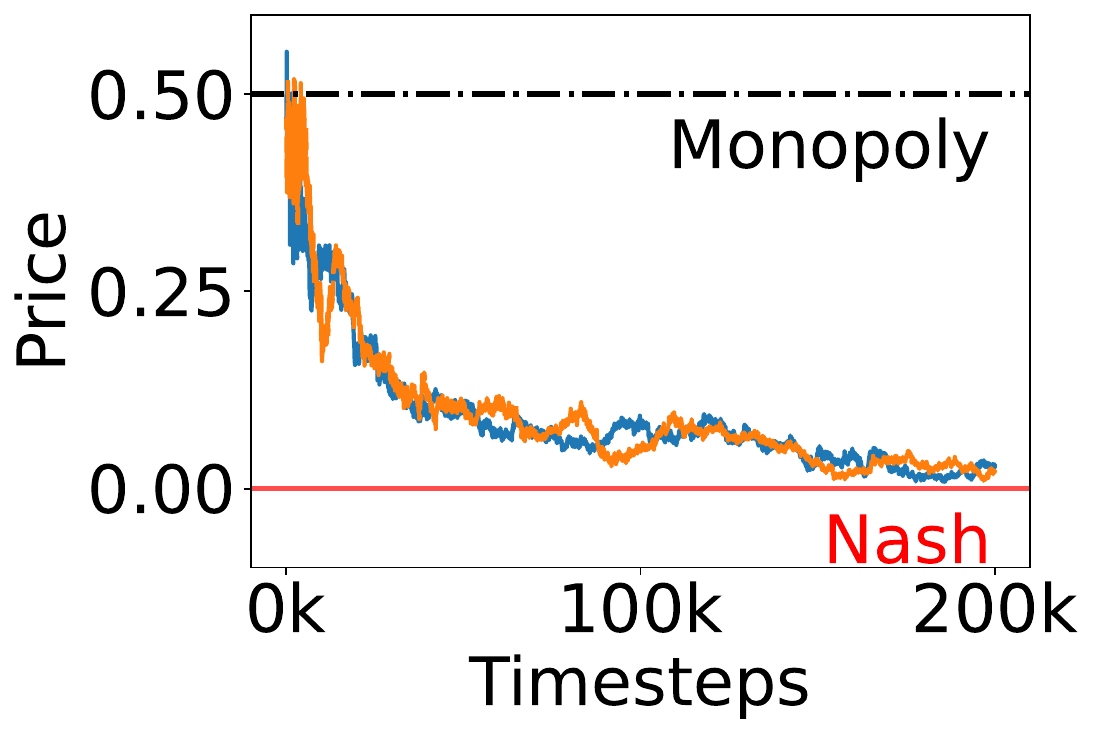}
    \caption{Competition}
    \label{fig:competition}
  \end{subfigure}%
  \hspace{5pt} 
  \begin{subfigure}{0.32\linewidth}
    \centering
    \includegraphics[width=\linewidth]{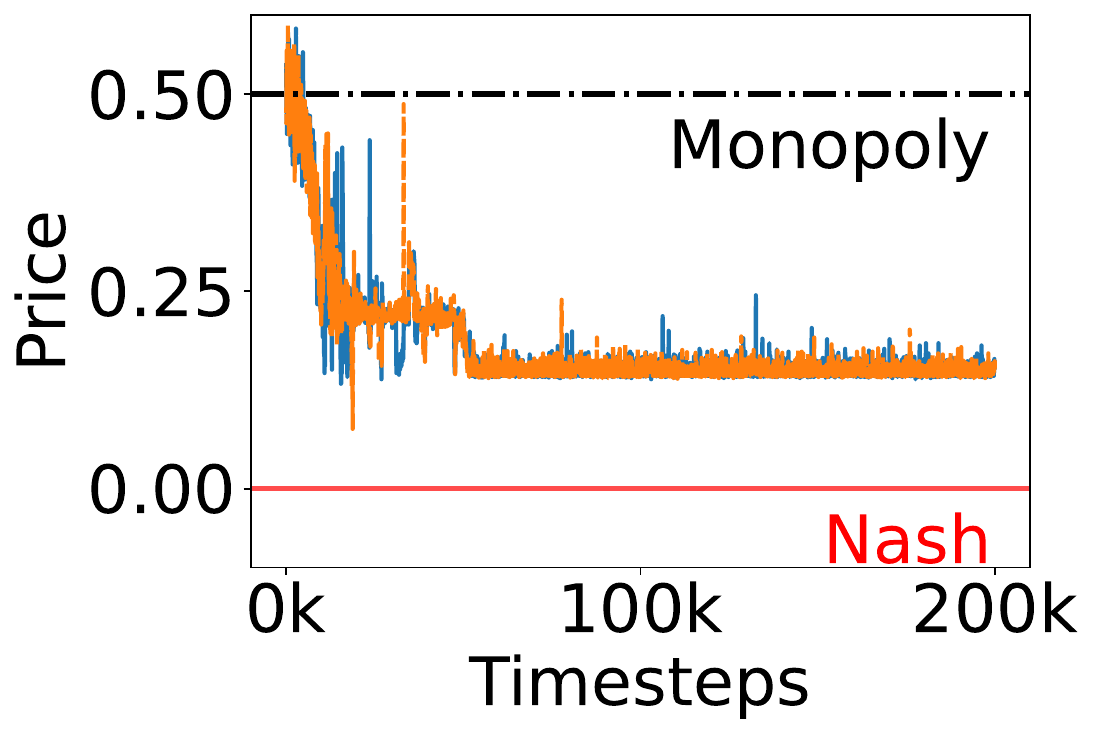}
    \caption{Collusion}
    \label{fig:collusion}
  \end{subfigure}%
  \hspace{5pt} 
  \begin{subfigure}{0.32\linewidth}
    \centering
    \includegraphics[width=\linewidth]{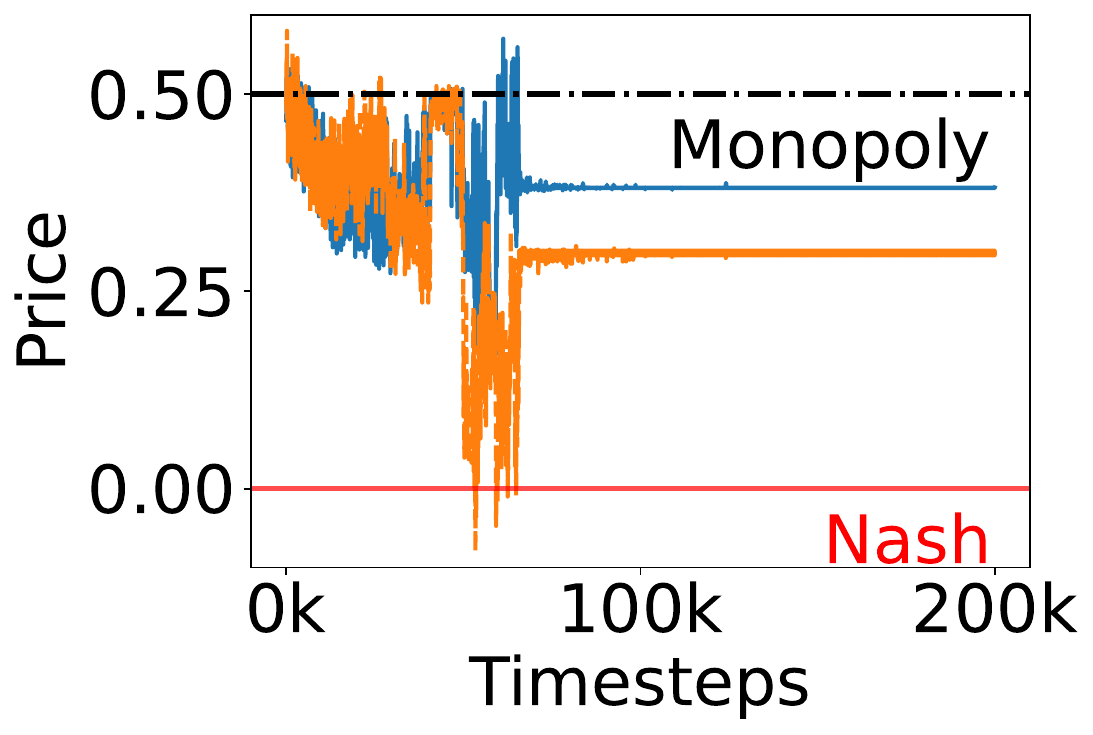}
    \caption{Dispersion}
    \label{fig:dispersion}
  \end{subfigure}
  
    \caption{Examples of pricing dynamics observed in our numerical study.}
    \label{fig:pricing_dynamics}
\end{figure}

In our experiments, we observe the three different pricing dynamics illustrated in Figure~\ref{fig:pricing_dynamics}. Figure~\ref{fig:competition} shows a competitive scenario: both agents' price choices stabilize around the Nash price, indicating a state of pure competition. Agents respond optimally to each other, with no opportunity to increase profits through deviation from their current strategy. Figure~\ref{fig:collusion} shows a collusive scenario: both agents' pricing strategies converge between the Nash and monopoly prices. The closer the pricing strategies are to the monopoly price, the higher the respective tacit collusion. Figure~\ref{fig:dispersion} shows a dispersive scenario: both agents converge to a stable but different price level, which leads to uneven profit margins between agents. We note that such an effect is unlikely to happen in practice and remains an artifact of specific algorithm's weaknesses as we will discuss in Section~\ref{subsec:discussion}. In the following, we discuss the respective effects and refer to Appendix~\ref{appendix:eta_kappa} for a detailed discussion on how we measure collusion and dispersion.


 Table~\ref{tab:algorithm_occurrences_combined} reports the distribution of competition (Comp.), collusion (Coll.), and dispersion (Disp.) for each algorithm and demand model over 50 seeds. Interestingly, the share of scenarios in which we observe each effect varies between the different algorithms and market models: first, it stands out that \gls{acr:tql} always leads to a collusive or dispersive dynamic but never reaches a competitive outcome. Second, we observe that the dynamics of each algorithm are sensitive to the respective demand model. The Bertrand Edgeworth model reinforces collusion significantly and suppresses competitive and dispersive outcomes across all algorithms, even for \gls{acr:tql}. Contrarily, the Bertrand Edgeworth model enables competitive outcomes for all \gls{acr:drl}-based algorithms, while only \gls{acr:ppo} with a continuous action space (PPO-C) yields competitive outcomes in the standard Bertrand model. Third, we note that \gls{acr:ppo} appears to be robust against dispersion dynamics over all studied demand models.

 \begin{table}[!hb]
\small
    \centering
\begin{tabular}{rrrrrrrrrr}
    \toprule
    & \multicolumn{3}{c}{Standard Bertrand} & \multicolumn{3}{c}{Bertrand Edgeworth} & \multicolumn{3}{c}{Logit Bertrand} \\
    \cmidrule(lr){2-4}\cmidrule(lr){5-7}\cmidrule(lr){8-10}
     \textbf{Alg.} & \textbf{Comp.} & \textbf{Coll.} & \textbf{Disp.} &  \textbf{Comp.} & \textbf{Coll.} & \textbf{Disp.} & \textbf{Comp.} & \textbf{Coll.} & \textbf{Disp.} \\
    \midrule
    \gls{acr:tql} & 0 \% & 68 \% & 32 \% & 0 \% & 94 \% & 6 \% & 0 \% & 52 \% & 48 \%  \\
    \gls{acr:dqn} & 0 \% & 36 \% & 64 \% & 0 \% & 100 \% & 0 \% & 38 \% & 62 \% & 0 \% \\
    \gls{acr:ppoc} & 46 \% & 54 \% & 0 \% & 0 \% & 100 \% & 0 \% & 76 \% & 24 \% & 0 \%\\
    \gls{acr:ppod} & 0 \% & 100 \% & 0 \% & 0 \% & 100 \% & 0 \% & 40 \% & 60 \% & 0 \%\\
    \gls{acr:sac} & 0 \% & 84 \% & 16 \% & 0 \% & 98 \% & 2 \% & 12 \% & 80 \% & 8 \% \\
    \bottomrule
\end{tabular}
    \caption{Distribution of pricing dynamics for each algorithm and market model over 50 seeds.}
    \label{tab:algorithm_occurrences_combined}
\end{table}

\begin{figure}[!t]
    \centering
    \begin{subfigure}{\textwidth}
        \centering
        \includegraphics[width=\linewidth]{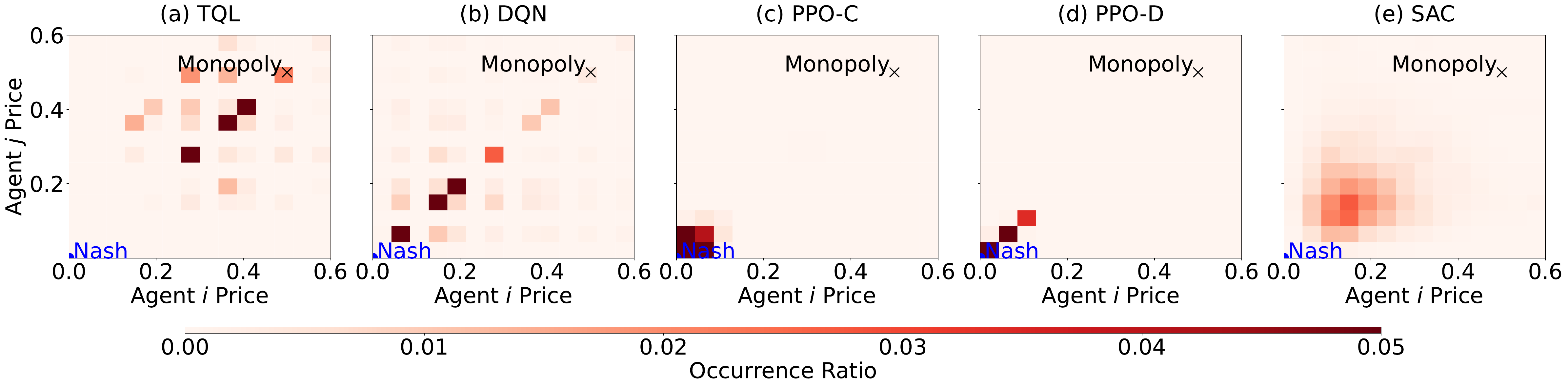}
        \caption{Standard Bertrand: states visited in last $10^4$ iterations.}
        \label{fig:standard_bertrand_price_heatmaps}
    \end{subfigure}
    \vfill 
    \begin{subfigure}{\textwidth}
        \centering
        \includegraphics[width=\linewidth]{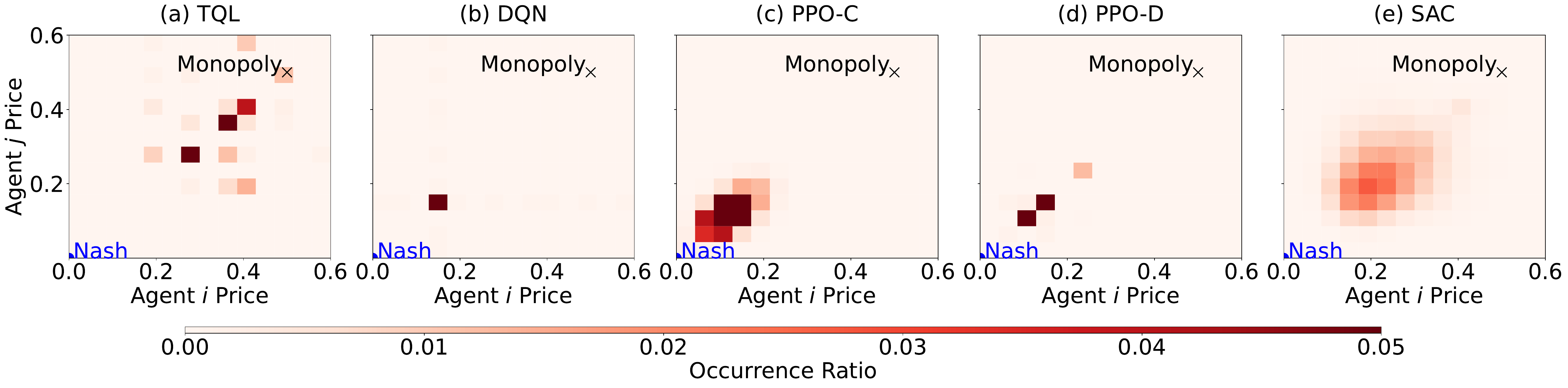}
        \caption{Bertrand Edgeworth: states visited in last $10^4$ iterations.}
        \label{fig:edgeworth_bertrand_price_heatmaps}
    \end{subfigure}
    \vfill 
    \begin{subfigure}{\textwidth}
        \centering
        \includegraphics[width=\linewidth]{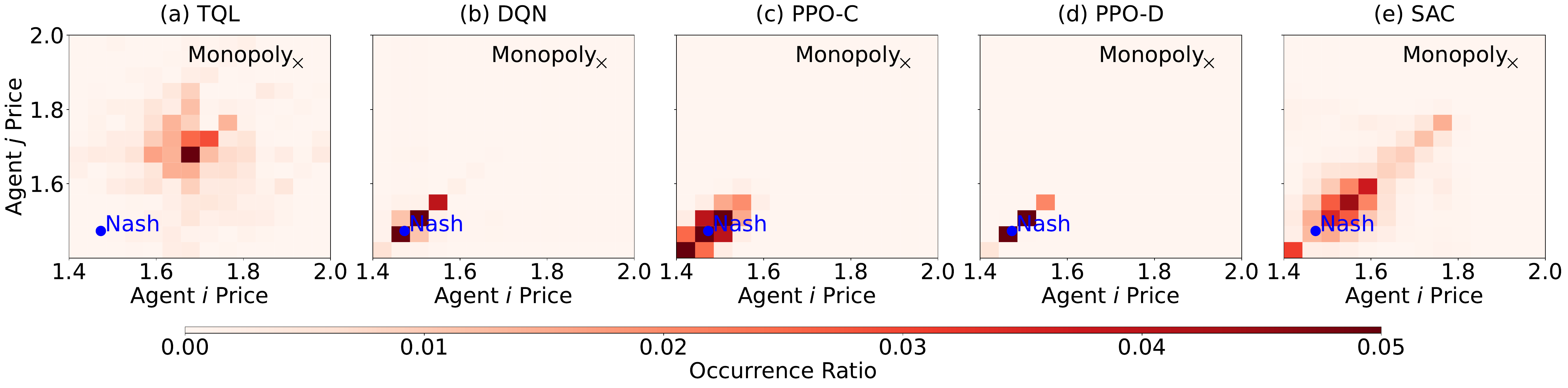}
        \caption{Logit Bertrand: states visited in last $10^4$ iterations.}
        \label{fig:logit_bertrand_price_heatmaps}
    \end{subfigure}
    \caption{Heatmaps of states visited in last $10^4$ iterations.}
    \label{fig:price_heatmaps}
\end{figure}

In addition to the frequency with which collusion and dispersion arise, it remains interesting to analyze how severe the respective collusion and dispersion are. Clearly, one aims to avoid any degree of collusion or dispersion, but from a practical perspective, algorithms colluding at a price level above but still close to the Nash price is less harmful than algorithms colluding at a price level close to the monopoly price. To visualize the pricing levels reached, Figure~\ref{fig:price_heatmaps} shows the price distribution between players for each algorithm and market model. These heatmaps can be interpreted in the following way: if our algorithms converge to a joint price level in a competitive or collusive way, the price distribution is distributed along the diagonal between the Nash and the monopoly price. While small deviations from this diagonal may arise due to numerical instability, large deviations indicate dispersion effects. As can be seen, the price levels on which we observe collusion or dispersion can vary significantly depending on the algorithm and market model. 

\begin{figure}[!b]
    \centering

    \begin{subfigure}{0.3\textwidth}
        \centering
\begin{tikzpicture}[scale=0.5]

\definecolor{darkgray176}{RGB}{176,176,176}
\small
\begin{axis}[
tick align=outside,
tick pos=left,
x grid style={darkgray176},
xmin=0.5, xmax=5.5,
xtick style={color=black},
xtick={1,2,3,4,5},
xticklabel style={rotate=45.0,anchor=east},
xticklabels={TQL,DQN,PPO-C,PPO-D,SAC},
y grid style={darkgray176},
ylabel={RPDI},
ymin=-0.2,
ymax=1.2,
ytick={-0.15,0,0.15,0.3,0.45,0.6,0.75,0.9,1.05,1.2},
ytick style={color=black}
]
\path [draw=black]
(axis cs:0.75,0.571428571428571)
--(axis cs:1.25,0.571428571428571)
--(axis cs:1.25,0.857142857142857)
--(axis cs:0.75,0.857142857142857)
--(axis cs:0.75,0.571428571428571)
--cycle;
\addplot [black]
table {%
1 0.571428571428571
1 0.285714285714286
};
\addplot [black]
table {%
1 0.857142857142857
1 1.14285714285714
};
\addplot [black]
table {%
0.875 0.285714285714286
1.125 0.285714285714286
};
\addplot [black]
table {%
0.875 1.14285714285714
1.125 1.14285714285714
};
\path [draw=black]
(axis cs:1.75,0.142857142857143)
--(axis cs:2.25,0.142857142857143)
--(axis cs:2.25,0.428571428571428)
--(axis cs:1.75,0.428571428571428)
--(axis cs:1.75,0.142857142857143)
--cycle;
\addplot [black]
table {%
2 0.142857142857143
2 0
};
\addplot [black]
table {%
2 0.428571428571428
2 0.714285714285714
};
\addplot [black]
table {%
1.875 0
2.125 0
};
\addplot [black]
table {%
1.875 0.714285714285714
2.125 0.714285714285714
};
\path [draw=black]
(axis cs:2.75,0.024764836)
--(axis cs:3.25,0.024764836)
--(axis cs:3.25,0.080629348)
--(axis cs:2.75,0.080629348)
--(axis cs:2.75,0.024764836)
--cycle;
\addplot [black]
table {%
3 0.024764836
3 0
};
\addplot [black]
table {%
3 0.080629348
3 0.16442579
};
\addplot [black]
table {%
2.875 0
3.125 0
};
\addplot [black]
table {%
2.875 0.16442579
3.125 0.16442579
};
\path [draw=black]
(axis cs:3.75,0.0857142857142856)
--(axis cs:4.25,0.0857142857142856)
--(axis cs:4.25,0.171428571428571)
--(axis cs:3.75,0.171428571428571)
--(axis cs:3.75,0.0857142857142856)
--cycle;
\addplot [black]
table {%
4 0.0857142857142856
4 0
};
\addplot [black]
table {%
4 0.171428571428571
4 0.257142857142857
};
\addplot [black]
table {%
3.875 0
4.125 0
};
\addplot [black]
table {%
3.875 0.257142857142857
4.125 0.257142857142857
};
\path [draw=black]
(axis cs:4.75,0.239624215)
--(axis cs:5.25,0.239624215)
--(axis cs:5.25,0.475704415)
--(axis cs:4.75,0.475704415)
--(axis cs:4.75,0.239624215)
--cycle;
\addplot [black]
table {%
5 0.239624215
5 1.205206e-05
};
\addplot [black]
table {%
5 0.475704415
5 0.829796
};
\addplot [black]
table {%
4.875 1.205206e-05
5.125 1.205206e-05
};
\addplot [black]
table {%
4.875 0.829796
5.125 0.829796
};
\addplot [very thick, black]
table {%
0.75 0.714285714285714
1.25 0.714285714285714
};
\addplot [very thick, black]
table {%
1.75 0.285714285714286
2.25 0.285714285714286
};
\addplot [very thick, black]
table {%
2.75 0.051604032
3.25 0.051604032
};
\addplot [very thick, black]
table {%
3.75 0.171428571428571
4.25 0.171428571428571
};
\addplot [very thick, black]
table {%
4.75 0.3353107
5.25 0.3353107
};
\end{axis}

\end{tikzpicture}
        \vspace{-0.3cm}
        \caption{Standard Bertrand}
        \label{fig:L10k_standard_bertrand_rpdi_boxplot1}
    \end{subfigure}%
    \hspace{0.02\textwidth}
    \begin{subfigure}{0.3\textwidth}
        \centering
\begin{tikzpicture}[scale=0.5]

\definecolor{darkgray176}{RGB}{176,176,176}
\small
\begin{axis}[
tick align=outside,
tick pos=left,
x grid style={darkgray176},
xmin=0.5, xmax=5.5,
xtick style={color=black},
xtick={1,2,3,4,5},
xticklabel style={rotate=45.0,anchor=east},
xticklabels={TQL,DQN,PPO-C,PPO-D,SAC},
y grid style={darkgray176},
ylabel={RPDI},
ymin=-0.2,
ymax=1.2,
ytick={-0.15,0,0.15,0.3,0.45,0.6,0.75,0.9,1.05,1.2},
ytick style={color=black}
]
\path [draw=black]
(axis cs:0.75,0.571428571428571)
--(axis cs:1.25,0.571428571428571)
--(axis cs:1.25,0.714285714285714)
--(axis cs:0.75,0.714285714285714)
--(axis cs:0.75,0.571428571428571)
--cycle;
\addplot [black]
table {%
1 0.571428571428571
1 0.428571428571428
};
\addplot [black]
table {%
1 0.714285714285714
1 0.857142857142857
};
\addplot [black]
table {%
0.875 0.428571428571428
1.125 0.428571428571428
};
\addplot [black]
table {%
0.875 0.857142857142857
1.125 0.857142857142857
};
\path [draw=black]
(axis cs:1.75,0.285714285714286)
--(axis cs:2.25,0.285714285714286)
--(axis cs:2.25,0.285714285714286)
--(axis cs:1.75,0.285714285714286)
--(axis cs:1.75,0.285714285714286)
--cycle;
\addplot [black]
table {%
2 0.285714285714286
2 0.285714285714286
};
\addplot [black]
table {%
2 0.285714285714286
2 0.285714285714286
};
\addplot [black]
table {%
1.875 0.285714285714286
2.125 0.285714285714286
};
\addplot [black]
table {%
1.875 0.285714285714286
2.125 0.285714285714286
};
\path [draw=black]
(axis cs:2.75,0.18617621)
--(axis cs:3.25,0.18617621)
--(axis cs:3.25,0.28015458)
--(axis cs:2.75,0.28015458)
--(axis cs:2.75,0.18617621)
--cycle;
\addplot [black]
table {%
3 0.18617621
3 0.045212806
};
\addplot [black]
table {%
3 0.28015458
3 0.42112172
};
\addplot [black]
table {%
2.875 0.045212806
3.125 0.045212806
};
\addplot [black]
table {%
2.875 0.42112172
3.125 0.42112172
};
\path [draw=black]
(axis cs:3.75,0.257142857142857)
--(axis cs:4.25,0.257142857142857)
--(axis cs:4.25,0.342857142857143)
--(axis cs:3.75,0.342857142857143)
--(axis cs:3.75,0.257142857142857)
--cycle;
\addplot [black]
table {%
4 0.257142857142857
4 0.171428571428571
};
\addplot [black]
table {%
4 0.342857142857143
4 0.428571428571429
};
\addplot [black]
table {%
3.875 0.171428571428571
4.125 0.171428571428571
};
\addplot [black]
table {%
3.875 0.428571428571429
4.125 0.428571428571429
};
\path [draw=black]
(axis cs:4.75,0.353829545)
--(axis cs:5.25,0.353829545)
--(axis cs:5.25,0.57056229)
--(axis cs:4.75,0.57056229)
--(axis cs:4.75,0.353829545)
--cycle;
\addplot [black]
table {%
5 0.353829545
5 0.028758645
};
\addplot [black]
table {%
5 0.57056229
5 0.89564574
};
\addplot [black]
table {%
4.875 0.028758645
5.125 0.028758645
};
\addplot [black]
table {%
4.875 0.89564574
5.125 0.89564574
};
\addplot [very thick, black]
table {%
0.75 0.571428571428571
1.25 0.571428571428571
};
\addplot [very thick, black]
table {%
1.75 0.285714285714286
2.25 0.285714285714286
};
\addplot [very thick, black]
table {%
2.75 0.23076928
3.25 0.23076928
};
\addplot [very thick, black]
table {%
3.75 0.257142857142857
4.25 0.257142857142857
};
\addplot [very thick, black]
table {%
4.75 0.45057268
5.25 0.45057268
};
\end{axis}

\end{tikzpicture}
        \vspace{-0.3cm}
        \caption{Bertrand Edgeworth}
        \label{fig:L10k_edgeworth_bertrand_rpdi_boxplot2}
    \end{subfigure}%
    \hspace{0.02\textwidth}
    \begin{subfigure}{0.3\textwidth}
        \centering
\begin{tikzpicture}[scale=0.5]

\definecolor{darkgray176}{RGB}{176,176,176}
\small
\begin{axis}[
tick align=outside,
tick pos=left,
x grid style={darkgray176},
xmin=0.5, xmax=5.5,
xtick style={color=black},
xtick={1,2,3,4,5},
xticklabel style={rotate=45.0,anchor=east},
xticklabels={TQL,DQN,PPO-C,PPO-D,SAC},
y grid style={darkgray176},
ylabel={RPDI},
ymin=-0.2,
ymax=1.2,
ytick={-0.15,0,0.15,0.3,0.45,0.6,0.75,0.9,1.05,1.2},
ytick style={color=black}
]
\path [draw=black]
(axis cs:0.75,0.328671017312842)
--(axis cs:1.25,0.328671017312842)
--(axis cs:1.25,0.585845653356134)
--(axis cs:0.75,0.585845653356134)
--(axis cs:0.75,0.328671017312842)
--cycle;
\addplot [black]
table {%
1 0.328671017312842
1 -0.0142284974115491
};
\addplot [black]
table {%
1 0.585845653356134
1 0.928745168080524
};
\addplot [black]
table {%
0.875 -0.0142284974115491
1.125 -0.0142284974115491
};
\addplot [black]
table {%
0.875 0.928745168080524
1.125 0.928745168080524
};
\path [draw=black]
(axis cs:1.75,-0.0142284974073175)
--(axis cs:2.25,-0.0142284974073175)
--(axis cs:2.25,0.0714963812695492)
--(axis cs:1.75,0.0714963812695492)
--(axis cs:1.75,-0.0142284974073175)
--cycle;
\addplot [black]
table {%
2 -0.0142284974073175
2 -0.0999533760926464
};
\addplot [black]
table {%
2 0.0714963812695492
2 0.157221259950646
};
\addplot [black]
table {%
1.875 -0.0999533760926464
2.125 -0.0999533760926464
};
\addplot [black]
table {%
1.875 0.157221259950646
2.125 0.157221259950646
};
\path [draw=black]
(axis cs:2.75,-0.0836207411504427)
--(axis cs:3.25,-0.0836207411504427)
--(axis cs:3.25,0.0555498340707963)
--(axis cs:2.75,0.0555498340707963)
--(axis cs:2.75,-0.0836207411504427)
--cycle;
\addplot [black]
table {%
3 -0.0836207411504427
3 -0.099953539823009
};
\addplot [black]
table {%
3 0.0555498340707963
3 0.264271238938053
};
\addplot [black]
table {%
2.875 -0.099953539823009
3.125 -0.099953539823009
};
\addplot [black]
table {%
2.875 0.264271238938053
3.125 0.264271238938053
};
\path [draw=black]
(axis cs:3.75,-0.0142284974115491)
--(axis cs:4.25,-0.0142284974115491)
--(axis cs:4.25,0.0714963812695492)
--(axis cs:3.75,0.0714963812695492)
--(axis cs:3.75,-0.0142284974115491)
--cycle;
\addplot [black]
table {%
4 -0.0142284974115491
4 -0.0999533760926464
};
\addplot [black]
table {%
4 0.0714963812695492
4 0.157221259950646
};
\addplot [black]
table {%
3.875 -0.0999533760926464
4.125 -0.0999533760926464
};
\addplot [black]
table {%
3.875 0.157221259950646
4.125 0.157221259950646
};
\path [draw=black]
(axis cs:4.75,0.0697819690265485)
--(axis cs:5.25,0.0697819690265485)
--(axis cs:5.25,0.304778263274336)
--(axis cs:4.75,0.304778263274336)
--(axis cs:4.75,0.0697819690265485)
--cycle;
\addplot [black]
table {%
5 0.0697819690265485
5 -0.099953539823009
};
\addplot [black]
table {%
5 0.304778263274336
5 0.657269247787611
};
\addplot [black]
table {%
4.875 -0.099953539823009
5.125 -0.099953539823009
};
\addplot [black]
table {%
4.875 0.657269247787611
5.125 0.657269247787611
};
\addplot [very thick, black]
table {%
0.75 0.500120774675036
1.25 0.500120774675036
};
\addplot [very thick, black]
table {%
1.75 0.0714963812483925
2.25 0.0714963812483925
};
\addplot [very thick, black]
table {%
2.75 -0.0104150442477879
3.25 -0.0104150442477879
};
\addplot [very thick, black]
table {%
3.75 0.0714963812695492
4.25 0.0714963812695492
};
\addplot [very thick, black]
table {%
4.75 0.177623451327434
5.25 0.177623451327434
};
\end{axis}

\end{tikzpicture}
        \vspace{-0.3cm}
        \caption{Logit Bertrand}       \label{fig:L10k_logit_bertrand_rpdi_boxplot3}
    \end{subfigure}
\caption{RPDI distribution for each algorithm and market model over the last $10^4$ iterations of each run.} 
\label{fig:rpdi_boxplots}
\end{figure}
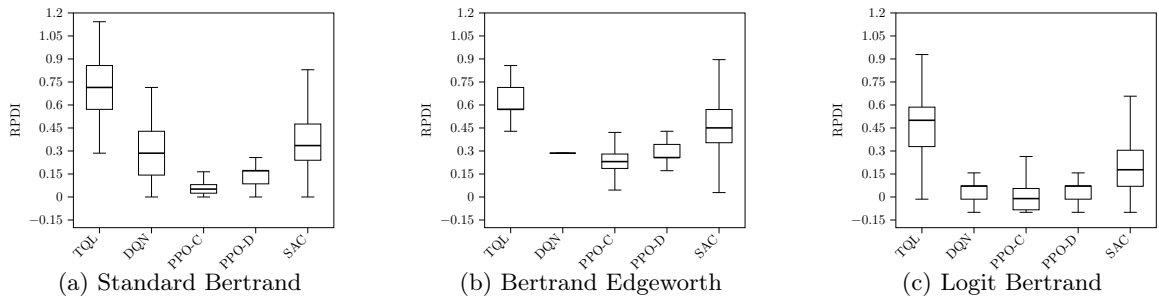

We note that not all dispersion effects are visible in Figure~\ref{fig:price_heatmaps}, because the occurrence ratio of dispersed prices often remains in the lower regime of the scale. Accordingly, we provide an additional price heatmap that uses a logarithmic scale in Figure~\ref{fig:price_heatmapsLOG} in Appendix~\ref{appendix:price_heatmaps_log}. Here, we focus on Figure~\ref{fig:price_heatmaps} to elucidate collusion dynamics, which we further discuss in the following.

To thoroughly discuss the degree of collusion beyond the visualization in Figure~\ref{fig:price_heatmaps}, we show the \gls{acr:rpdi} distribution (cf. Appendix~\ref{appendix:collusion_indicators}) for each algorithm and marked model in Figure~\ref{fig:rpdi_boxplots}. Here, a higher \gls{acr:rpdi} indicates a higher degree of collusion, i.e., that both algorithms collude on a higher price level, with an \gls{acr:rpdi} of one indicating collusion at the monopoly price. Similar to the variations in frequency, we see varying effects across algorithms and market models when analyzing the degree of collusion. First, we observe that the degree of collusion depends on the market model: all algorithms show the lowest degree of collusion in the Logit Bertrand model. Comparing the standard and Bertrand Edgeworth model, the effects are more nuanced. While both \gls{acr:ppo} and the \gls{acr:sac} algorithm show a lower degree of collusion in the standard Bertrand model, \gls{acr:dqn} shows a similar median in both models but a significantly smaller, almost nonexisting variance, in the Bertrand Edgeworth model. \gls{acr:tql} shows the highest degree of collusion and a large variance across all market models.

\subsection{Discussion}\label{subsec:discussion}
While our results remain primarily numerical, some of the effects that we observed can be explained by analyzing the studied algorithms and market models. 

\paragraph{Dispersion dynamics:} One may wonder why a dispersion effect that has also been observed in other studies in the \gls{acr:tql} case \citep[cf.][]{calvano2020artificial} arises at all, as it is irrational from an economic perspective. The reason for dispersion effects can be found in the nature of the studied algorithms: both \gls{acr:tql} and \gls{acr:dqn} are off-policy algorithms that require an explicit exploration mechanism to avoid getting stuck in local optima. Whenever dispersion arises, these algorithms get stuck in such local optima due to malfunctioning exploration. Contrarily, \gls{acr:ppo} is an on-policy algorithm that ensures exploration implicitly by directly parameterizing and learning from a gradually improving policy. Accordingly, the effects of malfunctioning exploration are less pronounced such that the algorithm is not prone to dispersion effects. Note that dispersion effects remain artificial: in practice, one would manually adjust an algorithm's outcome (or replace it at all) when observing dispersion.

\begin{figure}[!hb]
    \centering
    \begin{subfigure}[b]{0.3\textwidth}
        \includegraphics[width=\textwidth]{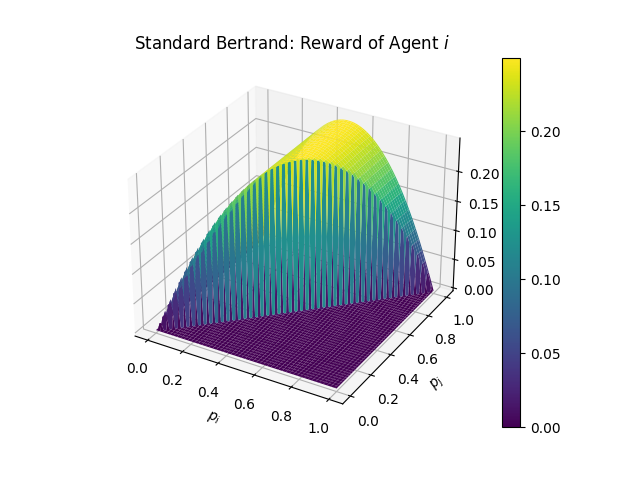}
        \caption{Standard Bertrand}
        \label{fig:img1}
    \end{subfigure}
    \hfill
    \begin{subfigure}[b]{0.3\textwidth}
        \includegraphics[width=\textwidth]{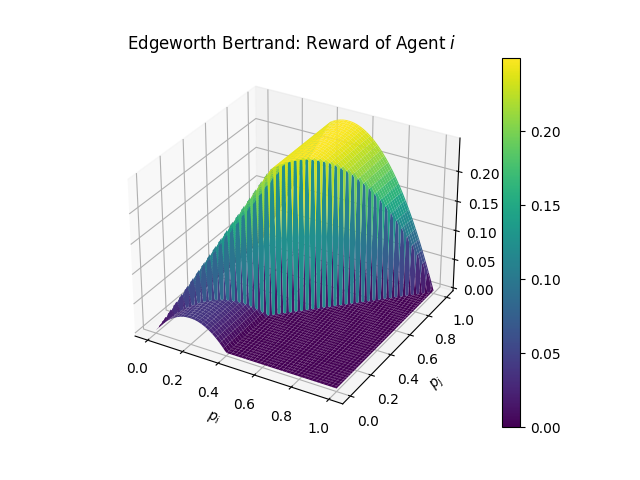}
        \caption{Bertrand Edgeworth}
        \label{fig:img2}
    \end{subfigure}
    \hfill
    \begin{subfigure}[b]{0.3\textwidth}
        \includegraphics[width=\textwidth]{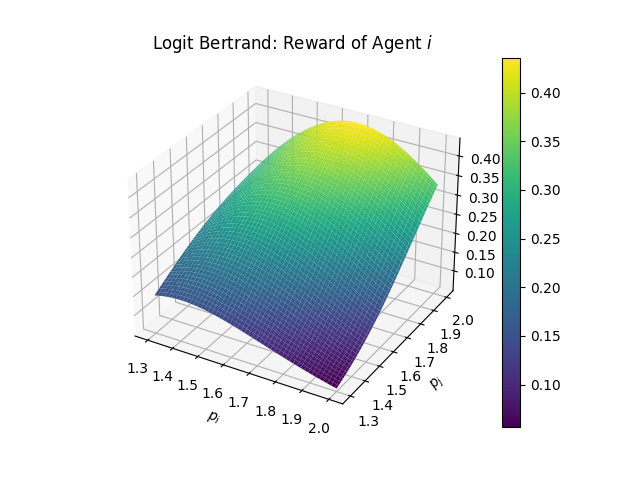}
        \caption{Logit Bertrand}
        \label{fig:img3}
    \end{subfigure}
    \caption{Reward of Player $i$ depending on the prices set by Player $i$ and Player $j$ for each market model.}
    \label{fig:rewardFunctions}
\end{figure}

\paragraph{Sensitivity to market models:} To understand the impact of the market model on the algorithms' pricing dynamics, we analyze the reward of Player $i$ with respect to the price set by Player $i$ and Player $j$ as shown in Figure~\ref{fig:rewardFunctions}. As can be seen, the characteristics of the reward functions differ with respect to their continuity and smoothness. As can be seen, the reward function in the Logit Bertrand model behaves better than the reward function in the standard Bertrand model, which again behaves better than the reward function in the Bertrand Edgeworth model. As the smoothness of the reward function can significantly ease or complicate the learning of a \gls{acr:drl} agent, it thus appears plausible that all algorithms but \gls{acr:tql} reach a competitive outcome in more scenarios when the market exhibits a better-behaved reward function.  

\paragraph{Superior performance of on-policy algorithms:} While we observe different pricing dynamics across market models and algorithms, the tested \gls{acr:ppo} variants show superior performance in all scenarios compared to the other algorithms tested: first, they never lead to dispersion effects; second, they most often reach competition; and third, in case of collusion, they collude at the lowest degrees. Here, the \gls{acr:ppo} variant with a continuous action space shows even better performance compared to its discrete action space counterpart. At first sight, it may feel counterintuitive that an algorithm that is usually seen as "data hungry" and sensitive to a huge amount of hyperparameters \citep{andrychowicz2020matters} can effectively mitigate collusion. However, once properly trained and parameterized \gls{acr:ppo} allows robust exploration implicitly by directly parameterizing and learning from a gradually improving policy. This allows to better escape local optima compared to the other algorithms analyzed and thus leads more often to a competitive consensus or decreased degrees of collusion.

\subsection{Limitations and future work}\label{subsec:limitations}
While our work sheds new light on algorithmic pricing collusion within the context of (deep) \gls{acr:rl}, several avenues for future research remain.

First, we limited our experiments to homogeneous settings in which only one type of algorithm competes with itself. While we made this design choice purposely to analyze the collusion potential of different algorithm types, we expect dynamics between a mixture of algorithms in practice. It will be interesting to see if a single on-policy algorithm that is less prone to collusion can dampen the collusive dynamics in a system with different competing algorithms. On the other hand, if companies understand that a particular algorithm leads to supra-competitive prices if they all use it, this might become the algorithm of choice. So, understanding a homogeneous setting is an important starting point.

Second, as in prior research, we limited our experiment to duopoly competition. While this decision allowed us to get unbiased insights into the respective dynamics without suffering from noise and convergence errors that can arise in multi-agent settings, it remains an interesting direction for future research to study settings with more than two agents. Here, it will be interesting to analyze if the presence of many agents reduces or amplifies the collusion effects we observe in the duopoly setting.

Third, when formulating our \gls{acr:mdp}, we assume that each agent observes both its own and its opponent's pricing decisions from the previous time step. Indeed, in many business-to-business markets, agents do not know the prices of their competitors. It will be an interesting avenue for future research to analyze if knowledge of past pricing decisions from other competing agents increases or decreases the risk of collusion.

Fourth, it will be interesting to focus detailed analyses on the initial and the final states of learning. Pricing agents on online platforms often change their prices multiple times per day. Nevertheless, there we cannot expect the same static setting for hundreds of thousands of rounds. It will be interesting to better understand the results after the convergence of the algorithms, but also in the initial phases. If we see a systematic bias and supra-competitive prices already in the initial stages, this will be more concerning as compared to situations where collusion is only learned after hundreds of thousands of rounds with static competition.

\section{Conclusion}
\label{sec:conclusion}
There is an ongoing debate about algorithmic collusion in pricing competition. Interestingly, most studies focus on \gls{acr:tql} and Bertrand competition with specific demand models. \gls{acr:drl} has received significant attention in academia and in business practice in recent years. Against this background, we ask the question of whether algorithmic collusion also arises with other types of algorithms, in particular \gls{acr:drl}. We provide a comprehensive analysis of tacit collusion in the context of algorithmic pricing via \gls{acr:rl}-based algorithms, focusing beyond \gls{acr:tql} on state-of-the-art  \gls{acr:drl} algorithms, studying a Bertrand competition in its standard variant but also a Bertrand-Edgeworth competition and a Bertrand competition with logit demand.

We present an extensive numerical study that partially confirms the findings of recent research but puts these into a different perspective: we confirm that algorithmic collusion is a concern with \gls{acr:tql} independent of the underlying market model but find different effects when employing \gls{acr:drl} algorithms: here, we observe collusion in all cases when analyzing dynamics in the Bertrand Edgeworth model, while the Logit Bertrand model leads to both scenarios in which we observe collusion and scenarios in which we observe competitive outcomes. In the standard Bertrand model, only one \gls{acr:ppo} variant reaches competitive outcomes. In general, our results suggest that collusion is impacted by an algorithm's exploration scheme, which is why on-policy algorithms that implicitly explore during learning lead to less collusion compared to the tested off-policy algorithms and \gls{acr:tql}.

Our work sheds light on the intricate dynamics of \gls{acr:drl} algorithms in pricing competitions. Our insights and the openly available implementation pave the way for further research, e.g., to analyze the dynamics of competition between a mixture of \gls{acr:drl} algorithms.

\singlespacing{
\bibliographystyle{model5-names}

\begin{thebibliography}{26}
\expandafter\ifx\csname natexlab\endcsname\relax\def\natexlab#1{#1}\fi
\providecommand{\url}[1]{\texttt{#1}}
\providecommand{\href}[2]{#2}
\providecommand{\path}[1]{#1}
\providecommand{\DOIprefix}{doi:}
\providecommand{\ArXivprefix}{arXiv:}
\providecommand{\URLprefix}{URL: }
\providecommand{\Pubmedprefix}{pmid:}
\providecommand{\doi}[1]{\href{http://dx.doi.org/#1}{\path{#1}}}
\providecommand{\Pubmed}[1]{\href{pmid:#1}{\path{#1}}}
\providecommand{\bibinfo}[2]{#2}
\ifx\xfnm\relax \def\xfnm[#1]{\unskip,\space#1}\fi
\bibitem[{Afshar et~al.(2022)Afshar, Rhuggenaath, Zhang \& Kaymak}]{afshar2022automated}
\bibinfo{author}{Afshar, R.~R.}, \bibinfo{author}{Rhuggenaath, J.}, \bibinfo{author}{Zhang, Y.}, \& \bibinfo{author}{Kaymak, U.} (\bibinfo{year}{2022}).
\newblock \bibinfo{title}{An automated deep reinforcement learning pipeline for dynamic pricing}.
\newblock {\itshape \bibinfo{journal}{IEEE Transactions on Artificial Intelligence}\/}, .
\bibitem[{Andrychowicz et~al.(2020)Andrychowicz, Raichuk, Sta{\'n}czyk, Orsini, Girgin, Marinier, Hussenot, Geist, Pietquin, Michalski et~al.}]{andrychowicz2020matters}
\bibinfo{author}{Andrychowicz, M.}, \bibinfo{author}{Raichuk, A.}, \bibinfo{author}{Sta{\'n}czyk, P.}, \bibinfo{author}{Orsini, M.}, \bibinfo{author}{Girgin, S.}, \bibinfo{author}{Marinier, R.}, \bibinfo{author}{Hussenot, L.}, \bibinfo{author}{Geist, M.}, \bibinfo{author}{Pietquin, O.}, \bibinfo{author}{Michalski, M.} et~al. (\bibinfo{year}{2020}).
\newblock \bibinfo{title}{What matters for on-policy deep actor-critic methods? a large-scale study}.
\newblock In {\itshape \bibinfo{booktitle}{International conference on learning representations}\/}.
\bibitem[{Asker et~al.(2021)Asker, Fershtman \& Pakes}]{asker2021artificial}
\bibinfo{author}{Asker, J.}, \bibinfo{author}{Fershtman, C.}, \& \bibinfo{author}{Pakes, A.} (\bibinfo{year}{2021}).
\newblock {\itshape \bibinfo{title}{Artificial intelligence and pricing: The impact of algorithm design}\/}.
\newblock \bibinfo{type}{Technical Report} National Bureau of Economic Research.
\bibitem[{Asker et~al.(2022)Asker, Fershtman \& Pakes}]{asker2022artificial}
\bibinfo{author}{Asker, J.}, \bibinfo{author}{Fershtman, C.}, \& \bibinfo{author}{Pakes, A.} (\bibinfo{year}{2022}).
\newblock \bibinfo{title}{Artificial intelligence, algorithm design, and pricing}.
\newblock In {\itshape \bibinfo{booktitle}{AEA Papers and Proceedings}\/} (pp. \bibinfo{pages}{452--56}).
\newblock volume \bibinfo{volume}{112}.
\bibitem[{Assad et~al.(2020)Assad, Clark, Ershov \& Xu}]{assad2020algorithmic}
\bibinfo{author}{Assad, S.}, \bibinfo{author}{Clark, R.}, \bibinfo{author}{Ershov, D.}, \& \bibinfo{author}{Xu, L.} (\bibinfo{year}{2020}).
\newblock \bibinfo{title}{Algorithmic pricing and competition: Empirical evidence from the german retail gasoline market}, .
\bibitem[{Bertrand(1883)}]{bertrand1883review}
\bibinfo{author}{Bertrand, J.} (\bibinfo{year}{1883}).
\newblock \bibinfo{title}{Review of “theorie mathematique de la richesse sociale” and of “recherches sur les principles mathematiques de la theorie des richesses.”}.
\newblock {\itshape \bibinfo{journal}{Journal de savants}\/},  {\itshape \bibinfo{volume}{67}\/}, \bibinfo{pages}{499}.
\bibitem[{Brown \& MacKay(2021)}]{brown2021competition}
\bibinfo{author}{Brown, Z.~Y.}, \& \bibinfo{author}{MacKay, A.} (\bibinfo{year}{2021}).
\newblock {\itshape \bibinfo{title}{Competition in pricing algorithms}\/}.
\newblock \bibinfo{type}{Technical Report} National Bureau of Economic Research.
\bibitem[{Calvano et~al.(2020)Calvano, Calzolari, Denicolo \& Pastorello}]{calvano2020artificial}
\bibinfo{author}{Calvano, E.}, \bibinfo{author}{Calzolari, G.}, \bibinfo{author}{Denicolo, V.}, \& \bibinfo{author}{Pastorello, S.} (\bibinfo{year}{2020}).
\newblock \bibinfo{title}{Artificial intelligence, algorithmic pricing, and collusion}.
\newblock {\itshape \bibinfo{journal}{American Economic Review}\/},  {\itshape \bibinfo{volume}{110}\/}, \bibinfo{pages}{3267--3297}.
\bibitem[{Capobianco \& Gonzaga(2020)}]{capobianco2020competition}
\bibinfo{author}{Capobianco, A.}, \& \bibinfo{author}{Gonzaga, P.} (\bibinfo{year}{2020}).
\newblock \bibinfo{title}{Competition challenges of big data: Algorithmic collusion, personalised pricing and privacy}.
\newblock In {\itshape \bibinfo{booktitle}{Legal Challenges of Big Data}\/} (pp. \bibinfo{pages}{46--63}).
\newblock \bibinfo{publisher}{Edward Elgar Publishing}.
\bibitem[{Charpentier et~al.(2021)Charpentier, Elie \& Remlinger}]{charpentier2021reinforcement}
\bibinfo{author}{Charpentier, A.}, \bibinfo{author}{Elie, R.}, \& \bibinfo{author}{Remlinger, C.} (\bibinfo{year}{2021}).
\newblock \bibinfo{title}{Reinforcement learning in economics and finance}.
\newblock {\itshape \bibinfo{journal}{Computational Economics}\/},  (pp. \bibinfo{pages}{1--38}).
\bibitem[{Constantine \& Quitaz(2018)}]{constantine2018oecd}
\bibinfo{author}{Constantine, S.}, \& \bibinfo{author}{Quitaz, V.} (\bibinfo{year}{2018}).
\newblock \bibinfo{title}{Oecd competition committee best practice roundtable--algorithms and collusion: United kingdom submission}.
\newblock {\itshape \bibinfo{journal}{Competition Law Journal}\/},  {\itshape \bibinfo{volume}{17}\/}, \bibinfo{pages}{41--48}.
\bibitem[{Edgeworth(1925)}]{edgeworth1925papers}
\bibinfo{author}{Edgeworth, F.~Y.} (\bibinfo{year}{1925}).
\newblock {\itshape \bibinfo{title}{Papers relating to political economy}\/} volume~\bibinfo{volume}{2}.
\newblock \bibinfo{publisher}{Royal Economic Society by Macmillan and Company, limited}.
\bibitem[{Epivent \& Lambin(2022)}]{epivent2022algorithmic}
\bibinfo{author}{Epivent, A.}, \& \bibinfo{author}{Lambin, X.} (\bibinfo{year}{2022}).
\newblock \bibinfo{title}{On algorithmic collusion and reward-punishment schemes}.
\newblock {\itshape \bibinfo{journal}{Available at SSRN 4227229}\/}, .
\bibitem[{Haarnoja et~al.(2018)Haarnoja, Zhou, Abbeel \& Levine}]{haarnoja2018soft}
\bibinfo{author}{Haarnoja, T.}, \bibinfo{author}{Zhou, A.}, \bibinfo{author}{Abbeel, P.}, \& \bibinfo{author}{Levine, S.} (\bibinfo{year}{2018}).
\newblock \bibinfo{title}{Soft actor-critic: Off-policy maximum entropy deep reinforcement learning with a stochastic actor}.
\newblock In {\itshape \bibinfo{booktitle}{International conference on machine learning}\/} (pp. \bibinfo{pages}{1861--1870}).
\newblock \bibinfo{organization}{PMLR}.
\bibitem[{Kastius \& Schlosser(2022)}]{kastius2022dynamic}
\bibinfo{author}{Kastius, A.}, \& \bibinfo{author}{Schlosser, R.} (\bibinfo{year}{2022}).
\newblock \bibinfo{title}{Dynamic pricing under competition using reinforcement learning}.
\newblock {\itshape \bibinfo{journal}{Journal of Revenue and Pricing Management}\/},  {\itshape \bibinfo{volume}{21}\/}, \bibinfo{pages}{50--63}.
\bibitem[{Klein(2021)}]{klein2021autonomous}
\bibinfo{author}{Klein, T.} (\bibinfo{year}{2021}).
\newblock \bibinfo{title}{Autonomous algorithmic collusion: Q-learning under sequential pricing}.
\newblock {\itshape \bibinfo{journal}{The RAND Journal of Economics}\/},  {\itshape \bibinfo{volume}{52}\/}, \bibinfo{pages}{538--558}.
\bibitem[{Liu et~al.(2019)Liu, Zhang, Wang, Deng \& Wu}]{liu2019dynamic}
\bibinfo{author}{Liu, J.}, \bibinfo{author}{Zhang, Y.}, \bibinfo{author}{Wang, X.}, \bibinfo{author}{Deng, Y.}, \& \bibinfo{author}{Wu, X.} (\bibinfo{year}{2019}).
\newblock \bibinfo{title}{Dynamic pricing on e-commerce platform with deep reinforcement learning: A field experiment}.
\newblock {\itshape \bibinfo{journal}{arXiv preprint arXiv:1912.02572}\/}, .
\bibitem[{Meylahn \& V.~den Boer(2022)}]{meylahn2022learning}
\bibinfo{author}{Meylahn, J.~M.}, \& \bibinfo{author}{V.~den Boer, A.} (\bibinfo{year}{2022}).
\newblock \bibinfo{title}{Learning to collude in a pricing duopoly}.
\newblock {\itshape \bibinfo{journal}{Manufacturing \& Service Operations Management}\/},  {\itshape \bibinfo{volume}{24}\/}, \bibinfo{pages}{2577--2594}.
\bibitem[{Mnih et~al.(2015)Mnih, Kavukcuoglu, Silver, Rusu, Veness, Bellemare, Graves, Riedmiller, Fidjeland, Ostrovski et~al.}]{mnih2015human}
\bibinfo{author}{Mnih, V.}, \bibinfo{author}{Kavukcuoglu, K.}, \bibinfo{author}{Silver, D.}, \bibinfo{author}{Rusu, A.~A.}, \bibinfo{author}{Veness, J.}, \bibinfo{author}{Bellemare, M.~G.}, \bibinfo{author}{Graves, A.}, \bibinfo{author}{Riedmiller, M.}, \bibinfo{author}{Fidjeland, A.~K.}, \bibinfo{author}{Ostrovski, G.} et~al. (\bibinfo{year}{2015}).
\newblock \bibinfo{title}{Human-level control through deep reinforcement learning}.
\newblock {\itshape \bibinfo{journal}{nature}\/},  {\itshape \bibinfo{volume}{518}\/}, \bibinfo{pages}{529--533}.
\bibitem[{Mosavi et~al.(2020)Mosavi, Faghan, Ghamisi, Duan, Ardabili, Salwana \& Band}]{mosavi2020comprehensive}
\bibinfo{author}{Mosavi, A.}, \bibinfo{author}{Faghan, Y.}, \bibinfo{author}{Ghamisi, P.}, \bibinfo{author}{Duan, P.}, \bibinfo{author}{Ardabili, S.~F.}, \bibinfo{author}{Salwana, E.}, \& \bibinfo{author}{Band, S.~S.} (\bibinfo{year}{2020}).
\newblock \bibinfo{title}{Comprehensive review of deep reinforcement learning methods and applications in economics}.
\newblock {\itshape \bibinfo{journal}{Mathematics}\/},  {\itshape \bibinfo{volume}{8}\/}, \bibinfo{pages}{1640}.
\bibitem[{Raffin et~al.(2021)Raffin, Hill, Gleave, Kanervisto, Ernestus \& Dormann}]{raffin2021stable}
\bibinfo{author}{Raffin, A.}, \bibinfo{author}{Hill, A.}, \bibinfo{author}{Gleave, A.}, \bibinfo{author}{Kanervisto, A.}, \bibinfo{author}{Ernestus, M.}, \& \bibinfo{author}{Dormann, N.} (\bibinfo{year}{2021}).
\newblock \bibinfo{title}{Stable-baselines3: Reliable reinforcement learning implementations}.
\newblock {\itshape \bibinfo{journal}{Journal of Machine Learning Research}\/},  {\itshape \bibinfo{volume}{22}\/}, \bibinfo{pages}{1--8}.
\bibitem[{Sanchez-Cartas \& Katsamakas(2022)}]{sanchez2022artificial}
\bibinfo{author}{Sanchez-Cartas, J.~M.}, \& \bibinfo{author}{Katsamakas, E.} (\bibinfo{year}{2022}).
\newblock \bibinfo{title}{Artificial intelligence, algorithmic competition and market structures}.
\newblock {\itshape \bibinfo{journal}{IEEE Access}\/},  {\itshape \bibinfo{volume}{10}\/}, \bibinfo{pages}{10575--10584}.
\bibitem[{Schulman et~al.(2017)Schulman, Wolski, Dhariwal, Radford \& Klimov}]{schulman2017proximal}
\bibinfo{author}{Schulman, J.}, \bibinfo{author}{Wolski, F.}, \bibinfo{author}{Dhariwal, P.}, \bibinfo{author}{Radford, A.}, \& \bibinfo{author}{Klimov, O.} (\bibinfo{year}{2017}).
\newblock \bibinfo{title}{Proximal policy optimization algorithms}.
\newblock {\itshape \bibinfo{journal}{arXiv preprint arXiv:1707.06347}\/}, .
\bibitem[{Watkins(1989)}]{watkins1989learning}
\bibinfo{author}{Watkins, C. J. C.~H.} (\bibinfo{year}{1989}).
\newblock \bibinfo{title}{Learning from delayed rewards}, .
\bibitem[{Werner(2023)}]{werner2023algorithmic}
\bibinfo{author}{Werner, T.} (\bibinfo{year}{2023}).
\newblock \bibinfo{title}{Algorithmic and human collusion}.
\newblock {\itshape \bibinfo{journal}{Available at SSRN 3960738}\/}, .
\bibitem[{Zhou et~al.(2022)Zhou, Yang \& Fu}]{zhou2022deep}
\bibinfo{author}{Zhou, Q.}, \bibinfo{author}{Yang, Y.}, \& \bibinfo{author}{Fu, S.} (\bibinfo{year}{2022}).
\newblock \bibinfo{title}{Deep reinforcement learning approach for solving joint pricing and inventory problem with reference price effects}.
\newblock {\itshape \bibinfo{journal}{Expert Systems with Applications}\/},  {\itshape \bibinfo{volume}{195}\/}, \bibinfo{pages}{116564}.

\end{thebibliography}

\newpage
\onehalfspacing
\begin{appendices}
	\normalsize
	

\section{}


\subsection{Hyperparameter tuning and experimental details}
\label{appendix:experimental_setup}

We used the same neural network architecture, a fully-connected feed-forward network with two hidden layers, each consisting of 256 nodes, for all \gls{acr:drl} algorithms. Such an architecture has been proven effective for various applications \citep{raffin2021stable}. While one can argue that it is possible to reach a better performance with individually tuned networks for each algorithm, we took this at first sight superficial design decision on purpose: using the same neural network for each algorithm allows us to focus our analyses on the collusion potential of each algorithm, independent of a potential bias that might arise when tuning each algorithm towards a varying neural network architecture.

We tuned the learning rate $\alpha$ and the discount factor $\gamma$, focusing on a Bertrand model with logit demand, fixing one agent's price at the Nash price and observing if and how fast the other agent converges to the Nash price. In this context, we tested learning rates $\alpha\in  [1e-5, 1e-4, 1e-3] $ and discount factors $\gamma \in  [0.95, 0.99, 0.999]$. We then selected an individual parameter configuration for each algorithm based on the fastest convergence observed over ten runs. Specifically, we set the discount factors $\gamma$ for \gls{acr:dqn}, \gls{acr:ppo}, and \gls{acr:sac} to 0.99; the learning rate $\alpha$ for \gls{acr:ppo} to  0.00005, for \gls{acr:dqn} to 0.0001, and for \gls{acr:sac} to 0.0003. 

For \gls{acr:tql}, we conducted separate parameter tuning for it. We selected the initial hyperparameter ranges from previous studies \citep{calvano2020artificial, klein2021autonomous} and iteratively adjusted these parameters, again choosing convergence speed as a selection criterion. We set the discount factor $\gamma$ for \gls{acr:tql} to 0.95 and the learning rate $\alpha$ to 0.125 to ensure fast and stable convergence.

Our experiments were conducted on high-performance computing equipment with an AMD Ryzen 9 7950X CPU (32 cores @ 4.5 GHz), 128 GB RAM, and an NVIDIA RTX 4090 GPU (24 GB). To enhance the credibility of our experiments, we ran each experiment 50 times independently and calculated the mean and standard deviation of the results to assess their stability and robustness. The code implements all algorithms and models used in our study, including the generation of figures in the results section.

\newpage

\subsection{LogNorm price heatmaps}
\label{appendix:price_heatmaps_log}

Figure \ref{fig:price_heatmaps} and Figure \ref{fig:price_heatmapsLOG} show the states visited in the last $10^4$ iterations for three different markets (Standard Bertrand, Bertrand Edgeworth, and Logit Bertrand) using linear and logarithmic Occurrence Ratio Bars respectively. Each market includes results for five algorithms (\gls{acr:tql}, \gls{acr:dqn}, \gls{acr:ppoc}, \gls{acr:ppod}, and \gls{acr:sac}). The linear scale is suitable for showing the absolute distribution of high-frequency states, while the logarithmic scale highlights the details of low-frequency states.

By observing both heatmaps, we can see the dispersion outcome for \gls{acr:tql}, \gls{acr:dqn}, and \gls{acr:sac} algorithms. In contrast, \gls{acr:ppo} algorithms show high aggregation - even when analyzing the logarithmic heatmap. 

\begin{figure}[ht]
    \centering
    \begin{subfigure}{\textwidth}
        \centering
        \includegraphics[width=\linewidth]{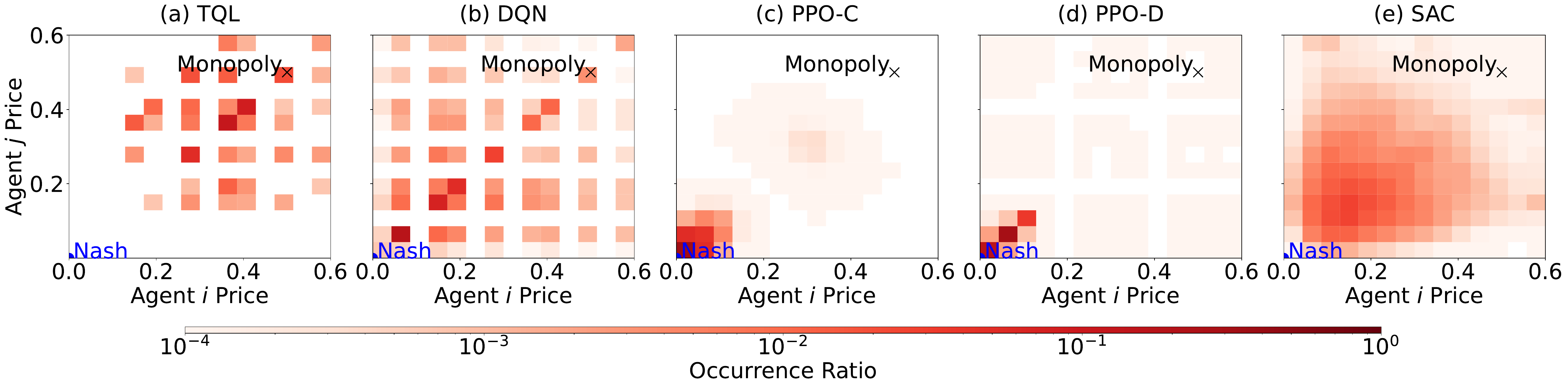}
        \caption{Standard Bertrand: states visited in last $10^4$ iterations.}
        \label{fig:standard_bertrand_price_heatmaps}
    \end{subfigure}
    \vfill 
    \begin{subfigure}{\textwidth}
        \centering
        \includegraphics[width=\linewidth]{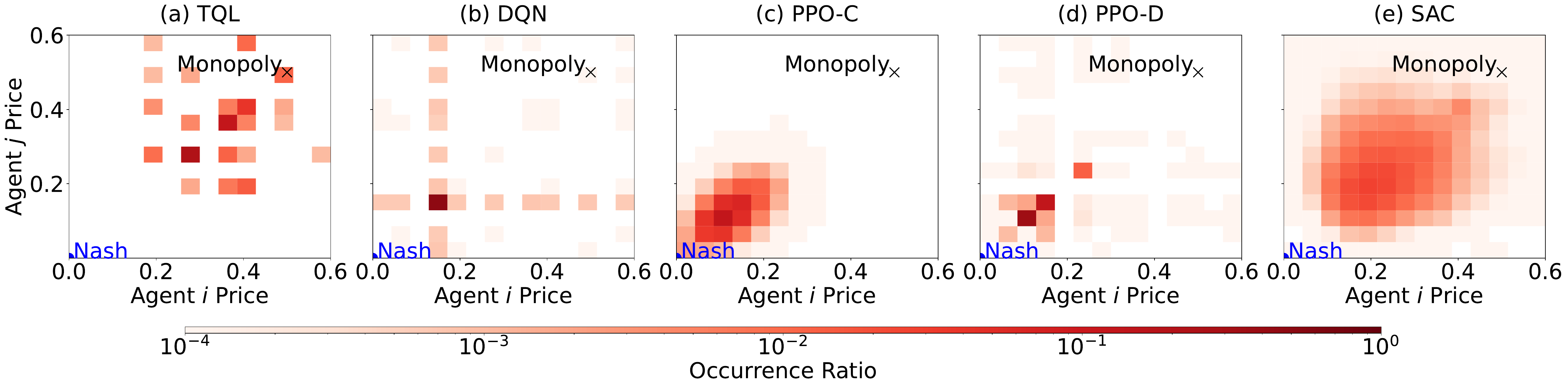}
        \caption{Bertrand Edgeworth: states visited in last $10^4$ iterations.}
        \label{fig:edgeworth_bertrand_price_heatmaps}
    \end{subfigure}
    \vfill 
    \begin{subfigure}{\textwidth}
        \centering
        \includegraphics[width=\linewidth]{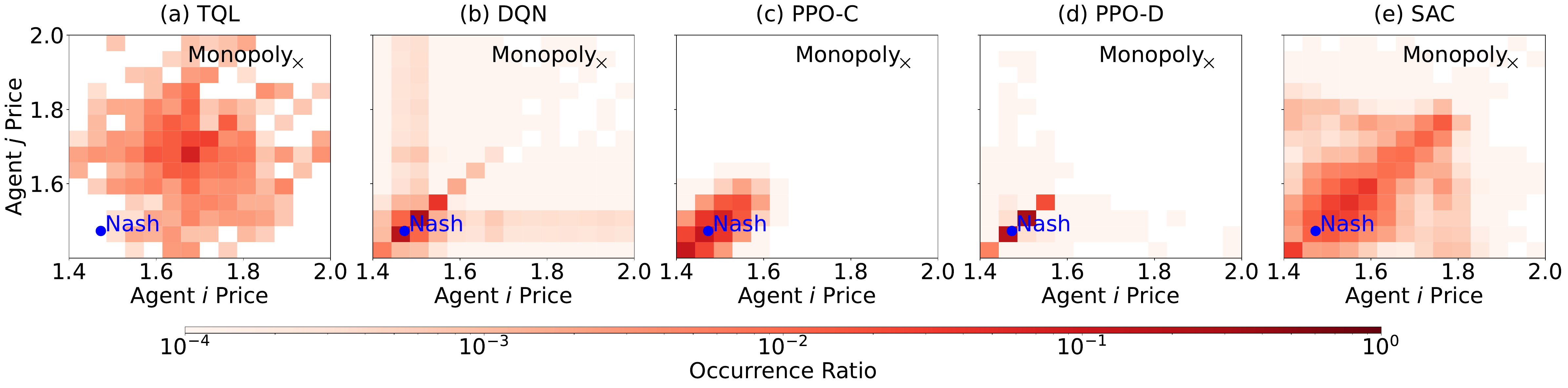}
        \caption{Logit Bertrand: states visited in last $10^4$ iterations.}
        \label{fig:logit_bertrand_price_heatmaps}
    \end{subfigure}
    \caption{LogNorm Occurrence Ratio Bar: Heatmaps of states visited in last $10^4$ iterations.}
    \label{fig:price_heatmapsLOG}
\end{figure}

\clearpage
\subsection{Description of quantitative indicators}
\label{appendix:eta_kappa}
Our analysis employs $\eta$ and $\kappa$ to assess pricing strategies' convergence, dispersion, and collusion levels between agents. The measure $\eta$ quantifies the degree of pricing convergence or dispersion, operationalized through the normalized average absolute price difference. Specifically, an $\eta$ from 0 to 0.2 reflects pricing strategies' convergence, and greater than 0.2 indicates the dispersion of the pricing strategies.
\begin{equation}
  \eta = \frac{1}{10000} \sum_{t=T-9999}^{T} \frac{|p^0_t - p^1_t|}{p^M - p^N}  
\end{equation}

High dispersion in pricing strategies implies that agents behave independently and lack coordination, making a collusion analysis superfluous. Hence, we only assess the level of collusion when the pricing strategies of the two agents exhibit convergence, i.e., $\eta < 0.2$. We introduce $\kappa$ as a measure to evaluate the level of price collusion. Its value is based on the normalized deviation of the average price set by two agents with similar pricing decisions from the Nash and monopoly prices. A $\kappa$ value close to 0 indicates competitive behavior, while a value close to 1 suggests increased collusion. We define $\kappa$ values below 0.05 as competitive behavior and those above 0.05 as indicative of collusion. This tolerance for deviations within the 0.05 range accounts for the noise potentially introduced during the learning and exploration processes of \gls{acr:rl} algorithms, as well as the possibility that the Nash price may not be accurately included in the discrete price space, allowing us to still consider such cases as approximating competitive scenarios.

\begin{equation}
    \kappa = \frac{1}{10000} \sum_{t=T-9999}^{T} \frac{p^0_t + p^1_t - 2p^N}{2(p^M - p^N)}
\end{equation}

\subsection{Collusion indicators}
\label{appendix:collusion_indicators}


In our study, we focus on each agent's individual pricing trends and profit trajectories to gain a more detailed understanding of their behavior. We examine two main indicators: \gls{acr:rpdi} and the profit metric $\Delta$ proposed by \citet{calvano2020artificial}. \gls{acr:rpdi} measures an agent's pricing relative to Nash equilibrium pricing and monopoly pricing, specifically indicating the extent to which an agent sets supra-competitive prices. The profit metric $\Delta$ assesses an agent's average profit over a period of time and is normalized relative to Nash and monopoly profits. Both indicators are calculated based on the values over the last 10,000 time steps to ensure the robustness and reliability of the analysis. 

Here, the \gls{acr:rpdi}  is defined as:
\begin{equation}
     \hat{p_i} = \frac{p_i - p^{\text{N}}}{p^{\text{M}} - p^{\text{N}}}
\end{equation}
which measures an agent's pricing relative to Nash and monopoly pricing. The profit metric proposed by \citet{calvano2020artificial} is calculated as:
\begin{equation}
     \Delta_i = \frac{\pi_i - \pi^N}{\pi^M - \pi^N}
\end{equation}
focusing on the change in each company $i$'s profits relative to Nash and monopoly profits. Together, these metrics paint a comprehensive picture: when the values of \gls{acr:rpdi}  and $\Delta_i$ for both agents are close to 0, it indicates behavior more akin to perfect competition; values near 1 suggest a higher propensity for collusion. Through this analysis, we not only reveal the relative pricing strategies and profit relationships among agents but also gain insight into each agent's independent pricing actions and profitability. 


\subsection{Further numerical results}

Moreover, to further illustrate our findings, we present the following tables and boxplots of the statistical measures for profit metric $\Delta$ and the \gls{acr:rpdi}. In Table \ref{tab:RPDI}, we present the mean and standard deviation of \gls{acr:rpdi} for various algorithms across different models. \gls{acr:tql} shows a high \gls{acr:rpdi} mean in both the standard Bertrand and Bertrand Edgeworth models, but slightly lower in the Logit Bertrand model. Furthermore, \gls{acr:dqn}'s\gls{acr:rpdi} mean ranges between 0.3 and 0.4 in the standard Bertrand and Bertrand Edgeworth models, but is much lower in the Logit Bertrand model, nearly close to 0. Notably, \gls{acr:ppoc} and \gls{acr:ppod} have relatively low \gls{acr:rpdi} means across all models, especially in the Logit Bertrand model, where \gls{acr:ppoc}'s mean is only 0.004, indicating more conservative pricing strategies. \gls{acr:sac}'s \gls{acr:rpdi} mean is close to \gls{acr:dqn}, falling between \gls{acr:dqn} and \gls{acr:tql} across all models.

Additionally, while both the \gls{acr:rpdi} and $\Delta$ values are normalized relative to Nash and monopoly prices and their corresponding profits, these two indicators are not perfectly correlated. This means that setting higher prices does not always guarantee higher profits. Therefore, it is necessary to use both price and profit metrics together to better understand the competitive landscape of the pricing market. Here, Figure \ref{fig:delta_boxplots} combines the $\Delta$ values of two agents over 50 experimental runs to visually represent the overall profit differences of each \gls{acr:rl} algorithm across different models. Moreover, Table \ref{tab:delta} provides more specific numerical results. By combining these two data sources, we can derive the following key observations: \gls{acr:tql} outperforms \gls{acr:sac}, \gls{acr:sac}  outperforms \gls{acr:dqn}, and \gls{acr:dqn}outperforms\gls{acr:ppod} and \gls{acr:ppoc}. These findings are consistent with our previous analysis.

Notably, \gls{acr:dqn} exhibits extremely low standard deviation (0.001) in the Bertrand Edgeworth model, indicating a high consistency in $\Delta$ values and demonstrating the stability of \gls{acr:dqn} in this model. In the Logit Bertrand model, the average profits of \gls{acr:dqn} and \gls{acr:ppo} are close to Nash profits, with correspondingly low standard deviations, indicating that their pricing strategies are cohesive and stable within the market structure. Furthermore, \gls{acr:sac}'s $\Delta$ standard deviation values across the three different Bertrand models are higher than those of the other four algorithms, suggesting that SAC has weaker cohesiveness and consistency. In general, compared to \gls{acr:tql}, \gls{acr:dqn}, and \gls{acr:sac}, \gls{acr:ppoc} and \gls{acr:ppod} consistently have lower average prices and standard deviations across different scenarios, highlighting the highly stable pricing decisions of the \gls{acr:ppo} algorithms.

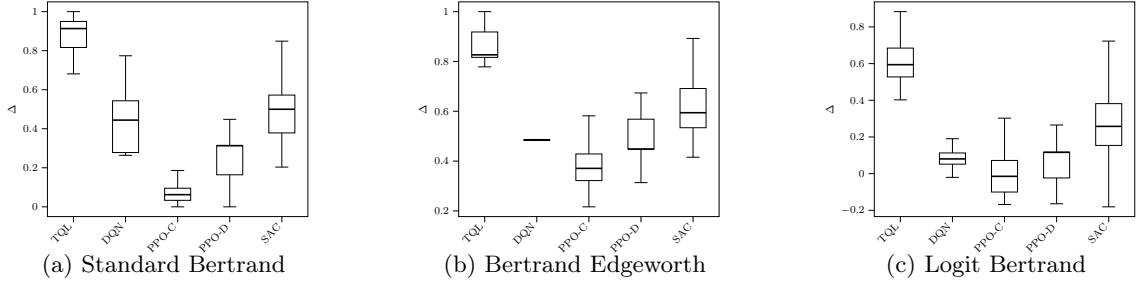
\begin{figure}[h]
    \centering

    \begin{subfigure}{0.3\textwidth}
        \centering
        \scriptsize
\begin{tikzpicture}[scale=0.5]

\definecolor{darkgray176}{RGB}{176,176,176}
\scriptsize

\begin{axis}[
tick align=outside,
tick pos=left,
x grid style={darkgray176},
xmin=0.5, xmax=5.5,
xtick style={color=black},
xtick={1,2,3,4,5},
xticklabel style={rotate=45.0,anchor=east},
xticklabels={TQL,DQN,PPO-C,PPO-D,SAC},
y grid style={darkgray176},
ylabel={\(\displaystyle \Delta\)},
ymin=-0.05, ymax=1.05,
ytick style={color=black}
]
\path [draw=black]
(axis cs:0.75,0.816326530612244)
--(axis cs:1.25,0.816326530612244)
--(axis cs:1.25,0.949616326530612)
--(axis cs:0.75,0.949616326530612)
--(axis cs:0.75,0.816326530612244)
--cycle;
\addplot [black]
table {%
1 0.816326530612244
1 0.681122448979592
};
\addplot [black]
table {%
1 0.949616326530612
1 1
};
\addplot [black]
table {%
0.875 0.681122448979592
1.125 0.681122448979592
};
\addplot [black]
table {%
0.875 1
1.125 1
};
\path [draw=black]
(axis cs:1.75,0.278201530612245)
--(axis cs:2.25,0.278201530612245)
--(axis cs:2.25,0.543264285714286)
--(axis cs:1.75,0.543264285714286)
--(axis cs:1.75,0.278201530612245)
--cycle;
\addplot [black]
table {%
2 0.278201530612245
2 0.263910204081632
};
\addplot [black]
table {%
2 0.543264285714286
2 0.773857142857143
};
\addplot [black]
table {%
1.875 0.263910204081632
2.125 0.263910204081632
};
\addplot [black]
table {%
1.875 0.773857142857143
2.125 0.773857142857143
};
\path [draw=black]
(axis cs:2.75,0.0332502226330128)
--(axis cs:3.25,0.0332502226330128)
--(axis cs:3.25,0.0952727635396727)
--(axis cs:2.75,0.0952727635396727)
--(axis cs:2.75,0.0332502226330128)
--cycle;
\addplot [black]
table {%
3 0.0332502226330128
3 0
};
\addplot [black]
table {%
3 0.0952727635396727
3 0.186166153874
};
\addplot [black]
table {%
2.875 0
3.125 0
};
\addplot [black]
table {%
2.875 0.186166153874
3.125 0.186166153874
};
\path [draw=black]
(axis cs:3.75,0.16409393877551)
--(axis cs:4.25,0.16409393877551)
--(axis cs:4.25,0.313469387755101)
--(axis cs:3.75,0.313469387755101)
--(axis cs:3.75,0.16409393877551)
--cycle;
\addplot [black]
table {%
4 0.16409393877551
4 0
};
\addplot [black]
table {%
4 0.313469387755101
4 0.448208081632653
};
\addplot [black]
table {%
3.875 0
4.125 0
};
\addplot [black]
table {%
3.875 0.448208081632653
4.125 0.448208081632653
};
\path [draw=black]
(axis cs:4.75,0.378677110920964)
--(axis cs:5.25,0.378677110920964)
--(axis cs:5.25,0.572447169748012)
--(axis cs:4.75,0.572447169748012)
--(axis cs:4.75,0.378677110920964)
--cycle;
\addplot [black]
table {%
5 0.378677110920964
5 0.203542664949654
};
\addplot [black]
table {%
5 0.572447169748012
5 0.848749903421702
};
\addplot [black]
table {%
4.875 0.203542664949654
5.125 0.203542664949654
};
\addplot [black]
table {%
4.875 0.848749903421702
5.125 0.848749903421702
};
\addplot [very thick, black]
table {%
0.75 0.913265306122449
1.25 0.913265306122449
};
\addplot [very thick, black]
table {%
1.75 0.444239795918367
2.25 0.444239795918367
};
\addplot [very thick, black]
table {%
2.75 0.0622889721236828
3.25 0.0622889721236828
};
\addplot [very thick, black]
table {%
3.75 0.313469387755101
4.25 0.313469387755101
};
\addplot [very thick, black]
table {%
4.75 0.499894732023981
5.25 0.499894732023981
};
\end{axis}

\end{tikzpicture}
        \vspace{-0.3cm}
        \caption{Standard Bertrand}
        \label{fig:L10k_standard_bertrand_delta_boxplot1}
    \end{subfigure}%
    \hspace{0.02\textwidth}
    \begin{subfigure}{0.3\textwidth}
        \centering
        \scriptsize
\begin{tikzpicture}[scale=0.5]

\definecolor{darkgray176}{RGB}{176,176,176}
\scriptsize

\begin{axis}[
tick align=outside,
tick pos=left,
x grid style={darkgray176},
xmin=0.5, xmax=5.5,
xtick style={color=black},
xtick={1,2,3,4,5},
xticklabel style={rotate=45.0,anchor=east},
xticklabels={TQL,DQN,PPO-C,PPO-D,SAC},
y grid style={darkgray176},
ylabel={\(\displaystyle \Delta\)},
ymin=0.177125873200128, ymax=1.03918448222857,
ytick style={color=black}
]
\path [draw=black]
(axis cs:0.75,0.816326530612244)
--(axis cs:1.25,0.816326530612244)
--(axis cs:1.25,0.918367346938775)
--(axis cs:0.75,0.918367346938775)
--(axis cs:0.75,0.816326530612244)
--cycle;
\addplot [black]
table {%
1 0.816326530612244
1 0.778279183673469
};
\addplot [black]
table {%
1 0.918367346938775
1 1
};
\addplot [black]
table {%
0.875 0.778279183673469
1.125 0.778279183673469
};
\addplot [black]
table {%
0.875 1
1.125 1
};
\path [draw=black]
(axis cs:1.75,0.484058367346938)
--(axis cs:2.25,0.484058367346938)
--(axis cs:2.25,0.485329795918367)
--(axis cs:1.75,0.485329795918367)
--(axis cs:1.75,0.484058367346938)
--cycle;
\addplot [black]
table {%
2 0.484058367346938
2 0.48313387755102
};
\addplot [black]
table {%
2 0.485329795918367
2 0.486924081632653
};
\addplot [black]
table {%
1.875 0.48313387755102
2.125 0.48313387755102
};
\addplot [black]
table {%
1.875 0.486924081632653
2.125 0.486924081632653
};
\path [draw=black]
(axis cs:2.75,0.32168581625461)
--(axis cs:3.25,0.32168581625461)
--(axis cs:3.25,0.428593782798657)
--(axis cs:2.75,0.428593782798657)
--(axis cs:2.75,0.32168581625461)
--cycle;
\addplot [black]
table {%
3 0.32168581625461
3 0.216310355428694
};
\addplot [black]
table {%
3 0.428593782798657
3 0.581679125960548
};
\addplot [black]
table {%
2.875 0.216310355428694
3.125 0.216310355428694
};
\addplot [black]
table {%
2.875 0.581679125960548
3.125 0.581679125960548
};
\path [draw=black]
(axis cs:3.75,0.448159591836735)
--(axis cs:4.25,0.448159591836735)
--(axis cs:4.25,0.56812787755102)
--(axis cs:3.75,0.56812787755102)
--(axis cs:3.75,0.448159591836735)
--cycle;
\addplot [black]
table {%
4 0.448159591836735
4 0.313450040816326
};
\addplot [black]
table {%
4 0.56812787755102
4 0.673469387755102
};
\addplot [black]
table {%
3.875 0.313450040816326
4.125 0.313450040816326
};
\addplot [black]
table {%
3.875 0.673469387755102
4.125 0.673469387755102
};
\path [draw=black]
(axis cs:4.75,0.533846690458206)
--(axis cs:5.25,0.533846690458206)
--(axis cs:5.25,0.691332952176306)
--(axis cs:4.75,0.691332952176306)
--(axis cs:4.75,0.533846690458206)
--cycle;
\addplot [black]
table {%
5 0.533846690458206
5 0.415422229690783
};
\addplot [black]
table {%
5 0.691332952176306
5 0.892457957327387
};
\addplot [black]
table {%
4.875 0.415422229690783
5.125 0.415422229690783
};
\addplot [black]
table {%
4.875 0.892457957327387
5.125 0.892457957327387
};
\addplot [very thick, black]
table {%
0.75 0.826530612244897
1.25 0.826530612244897
};
\addplot [very thick, black]
table {%
1.75 0.484549795918367
2.25 0.484549795918367
};
\addplot [very thick, black]
table {%
2.75 0.370652413839772
3.25 0.370652413839772
};
\addplot [very thick, black]
table {%
3.75 0.448163265306122
4.25 0.448163265306122
};
\addplot [very thick, black]
table {%
4.75 0.594092819795769
5.25 0.594092819795769
};
\end{axis}

\end{tikzpicture}
        \vspace{-0.3cm}
        \caption{Bertrand Edgeworth}
        \label{fig:L10k_edgeworth_bertrand_delta_boxplot2}
    \end{subfigure}%
    \hspace{0.02\textwidth}
    \begin{subfigure}{0.3\textwidth}
        \centering
        \scriptsize
\begin{tikzpicture}[scale=0.5]

\definecolor{darkgray176}{RGB}{176,176,176}
\scriptsize

\begin{axis}[
tick align=outside,
tick pos=left,
x grid style={darkgray176},
xmin=0.5, xmax=5.5,
xtick style={color=black},
xtick={1,2,3,4,5},
xticklabel style={rotate=45.0,anchor=east},
xticklabels={TQL,DQN,PPO-C,PPO-D,SAC},
y grid style={darkgray176},
ylabel={\(\displaystyle \Delta\)},
ymin=-0.234231997336179, ymax=0.936607658916398,
ytick style={color=black}
]
\path [draw=black]
(axis cs:0.75,0.527385126130069)
--(axis cs:1.25,0.527385126130069)
--(axis cs:1.25,0.6845450876111)
--(axis cs:0.75,0.6845450876111)
--(axis cs:0.75,0.527385126130069)
--cycle;
\addplot [black]
table {%
1 0.527385126130069
1 0.40227838137647
};
\addplot [black]
table {%
1 0.6845450876111
1 0.883387674541281
};
\addplot [black]
table {%
0.875 0.40227838137647
1.125 0.40227838137647
};
\addplot [black]
table {%
0.875 0.883387674541281
1.125 0.883387674541281
};
\path [draw=black]
(axis cs:1.75,0.0516762240969657)
--(axis cs:2.25,0.0516762240969657)
--(axis cs:2.25,0.11273131977691)
--(axis cs:1.75,0.11273131977691)
--(axis cs:1.75,0.0516762240969657)
--cycle;
\addplot [black]
table {%
2 0.0516762240969657
2 -0.0202843049803973
};
\addplot [black]
table {%
2 0.11273131977691
2 0.190282146491619
};
\addplot [black]
table {%
1.875 -0.0202843049803973
2.125 -0.0202843049803973
};
\addplot [black]
table {%
1.875 0.190282146491619
2.125 0.190282146491619
};
\path [draw=black]
(axis cs:2.75,-0.100365505722417)
--(axis cs:3.25,-0.100365505722417)
--(axis cs:3.25,0.0718790341541684)
--(axis cs:2.75,0.0718790341541684)
--(axis cs:2.75,-0.100365505722417)
--cycle;
\addplot [black]
table {%
3 -0.100365505722417
3 -0.168363262373528
};
\addplot [black]
table {%
3 0.0718790341541684
3 0.302486053298224
};
\addplot [black]
table {%
2.875 -0.168363262373528
3.125 -0.168363262373528
};
\addplot [black]
table {%
2.875 0.302486053298224
3.125 0.302486053298224
};
\path [draw=black]
(axis cs:3.75,-0.0235246931164227)
--(axis cs:4.25,-0.0235246931164227)
--(axis cs:4.25,0.116774190836423)
--(axis cs:3.75,0.116774190836423)
--(axis cs:3.75,-0.0235246931164227)
--cycle;
\addplot [black]
table {%
4 -0.0235246931164227
4 -0.164476396920077
};
\addplot [black]
table {%
4 0.116774190836423
4 0.265187989007534
};
\addplot [black]
table {%
3.875 -0.164476396920077
4.125 -0.164476396920077
};
\addplot [black]
table {%
3.875 0.265187989007534
4.125 0.265187989007534
};
\path [draw=black]
(axis cs:4.75,0.153826029486214)
--(axis cs:5.25,0.153826029486214)
--(axis cs:5.25,0.381753520982571)
--(axis cs:4.75,0.381753520982571)
--(axis cs:4.75,0.153826029486214)
--cycle;
\addplot [black]
table {%
5 0.153826029486214
5 -0.181012012961062
};
\addplot [black]
table {%
5 0.381753520982571
5 0.722671537578917
};
\addplot [black]
table {%
4.875 -0.181012012961062
5.125 -0.181012012961062
};
\addplot [black]
table {%
4.875 0.722671537578917
5.125 0.722671537578917
};
\addplot [very thick, black]
table {%
0.75 0.594124904345293
1.25 0.594124904345293
};
\addplot [very thick, black]
table {%
1.75 0.0802218385161605
2.25 0.0802218385161605
};
\addplot [very thick, black]
table {%
2.75 -0.0146671865768571
3.25 -0.0146671865768571
};
\addplot [very thick, black]
table {%
3.75 0.116178383196044
4.25 0.116178383196044
};
\addplot [very thick, black]
table {%
4.75 0.25766429998538
5.25 0.25766429998538
};
\end{axis}

\end{tikzpicture}
        \vspace{-0.3cm}
        \caption{Logit Bertrand}      \label{fig:L10k_logit_bertrand_delta_boxplot3}
    \end{subfigure}
\caption{$\Delta$ distribution for each algorithm and market model over the last $10^4$ iterations of each run.} 
\label{fig:delta_boxplots}
\end{figure}

\begin{table}[ht]
    \centering
    \setlength{\tabcolsep}{2pt} 
    \renewcommand{\arraystretch}{1.2} 
    \resizebox{0.8\columnwidth}{!}{
    
    \begin{tabular}{l l S[table-format=1.3] S[table-format=1.3] S[table-format=1.3] S[table-format=1.3] S[table-format=1.3] S[table-format=1.3]}
        \hline
         &  & \multicolumn{2}{c}{Standard Bertrand} & \multicolumn{2}{c}{Bertrand Edgeworth} & \multicolumn{2}{c}{Logit Bertrand} \\
         \cmidrule(lr){3-4}  \cmidrule(lr){5-6} \cmidrule(lr){7-8}
       Algorithm & Statistic & {$\hat{p}_0$} & {$\hat{p}_1$} & {$\hat{p}_0$} & {$\hat{p}_1$} & {$\hat{p}_0$} & {$\hat{p}_1$} \\   
        \hline
        \gls{acr:tql} & Mean & 0.782 & 0.868 & 0.823 & 0.826 & 0.481 & 0.491 \\
        & Std & 0.221 & 0.400 & 0.305 & 0.325 & 0.162 & 0.197 \\
        \hline
        \gls{acr:dqn} & Mean & 0.398 & 0.389 & 0.300 & 0.301 & 0.057 & 0.058 \\
        & Std & 0.366 & 0.349 & 0.131 & 0.135 & 0.101 & 0.102 \\
        \hline
        \gls{acr:ppoc} & Mean & 0.056 & 0.062 & 0.238 & 0.234 & 0.004 & 0.004 \\
        & Std & 0.044 & 0.070 & 0.074 & 0.070 & 0.080 & 0.083 \\
        \hline
        \gls{acr:ppod} & Mean & 0.151 & 0.148 & 0.284 & 0.285 & 0.042 & 0.052 \\
        & Std & 0.053 & 0.051 & 0.043 & 0.047 & 0.046 & 0.050 \\
        \hline
        \gls{acr:sac} & Mean & 0.382 & 0.387 & 0.472 & 0.472 & 0.208 & 0.207 \\
        & Std & 0.217 & 0.222 & 0.165 & 0.165 & 0.206 & 0.196 \\
        \hline
    \end{tabular}
    }
    \caption{Comparison of \gls{acr:rpdi} across models for different algorithms.}
    \label{tab:RPDI}
\end{table}

\begin{table}[ht]
    \centering
    \setlength{\tabcolsep}{2pt} 
    \renewcommand{\arraystretch}{1.2} 
    \resizebox{0.8\columnwidth}{!}{
    \begin{tabular}{l l S[table-format=1.3] S[table-format=1.3] S[table-format=1.3] S[table-format=1.3] S[table-format=1.3] S[table-format=1.3]}
        \hline
        & & \multicolumn{2}{c}{Standard Bertrand} & \multicolumn{2}{c}{Bertrand Edgeworth} & \multicolumn{2}{c}{Logit Bertrand} \\
        \cmidrule{3-4} \cmidrule{5-6} \cmidrule{7-8}
        Algorithm & Statistic & {$\Delta_0$} & {$\Delta_1$} & {$\Delta_0$} & {$\Delta_1$} & {$\Delta_0$} & {$\Delta_1$} \\ \hline
        \multirow{2}{*}{\gls{acr:tql}} & Mean & 0.891 & 0.880 & 0.883 & 0.882 & 0.665 & 0.628 \\
                             & Std  & 0.070 & 0.064 & 0.085 & 0.079 & 0.148 & 0.109 \\
        \hline
        \multirow{2}{*}{\gls{acr:dqn}} & Mean & 0.425 & 0.462 & 0.485 & 0.484 & 0.074 & 0.071 \\
                             & Std  & 0.113 & 0.144 & 0.001 & 0.001 & 0.057 & 0.049 \\
        \hline
        \multirow{2}{*}{\gls{acr:ppoc}} & Mean & 0.070 & 0.074 & 0.375 & 0.383 & -0.011 & -0.012 \\
                             & Std  & 0.051 & 0.051 & 0.069 & 0.078 & 0.123 & 0.116 \\
        \hline
        \multirow{2}{*}{\gls{acr:ppod}} & Mean & 0.258 & 0.273 & 0.481 & 0.478 & 0.076 & 0.057 \\
                             & Std  & 0.103 & 0.103 & 0.074 & 0.075 & 0.083 &  0.078 \\ 
        \hline
        \multirow{2}{*}{\gls{acr:sac}} & Mean & 0.489 & 0.488 & 0.629 & 0.614 & 0.286 & 0.293 \\
                             & Std  & 0.138 & 0.143 & 0.116 & 0.102 & 0.216 &  0.268 \\ \hline     
    \end{tabular}
    }
    \caption{Comparison of $\Delta$ across models for different algorithms.}
    \label{tab:delta}
\end{table}

\end{appendices}

\end{document}